\documentclass[11pt,a4paper]{article}
\pdfoutput=1
\usepackage{myjheppub}
\usepackage{latexsym,amsfonts,amsmath,amssymb,textgreek,upgreek}
\usepackage{amsthm}
\usepackage{mathtools}
\usepackage{bbm}
\usepackage{graphicx}
\usepackage{color}
\usepackage{amsfonts}
\usepackage{caption}
\usepackage{subcaption}
\usepackage[normalem]{ulem}
\usepackage{comment}
\usepackage{url}
\usepackage{slashed}
\usepackage{bm}
\usepackage{hhline}
\usepackage{physics}
\usepackage{booktabs}
\usepackage{siunitx}
\usepackage{tikz}
\usepackage{tabu}
\usepackage{diagbox} 
\usepackage{hyperref}
\usepackage{pdflscape}
\usepackage[utf8]{inputenc}
\usepackage[T1]{fontenc}

\newtheorem{conjecture}{Conjecture}[section]

\usepackage{footmisc}

\usepackage{tikz}
\usetikzlibrary{arrows}

\newcommand{\CN}{\mathcal{N}}

\newcommand{\PCH}{\text{PCH}}

\newcommand{\bd}{\boldsymbol}

\renewcommand{\Tr}{\mbox{Tr}}

\newcommand{\sgn}{\mbox{sgn}}

\newcommand{\IC}{\mathbb{C}}
\newcommand{\IZ}{\mathbb{Z}}

\newcommand{\IN}{\mathbb{N}}

\def\R{\mathbb{R}}

\newcommand\be{\begin{equation}}
\newcommand\ee{\end{equation}}
\newcommand\bea{\begin{eqnarray}}
\newcommand\eea{\end{eqnarray}}

\renewcommand{\=}{\;= \;}

\newcommand{\wt}{\widetilde}

\newcommand{\p}{\partial}

\newcommand{\ndt}{\noindent}

\newcommand{\blue}{\textcolor{blue}}

\newcommand{\RI}{$\frac14$}

\title{Fermionic trace relations and supersymmetric indices at finite~$N$}

\author{Giorgos Eleftheriou$^1$, Ziming Ji$^2$, Sameer Murthy$^1$}

\affiliation{${}^1$ Department of Mathematics, King's College London, The Strand, London WC2R 2LS, UK}
   
\affiliation{${}^2$  Department of Physics, Northeastern University, Boston, MA, 02115, USA}

\emailAdd{geleftheriou4@gmail.com}
\emailAdd{zi.ji@northeastern.edu}
\emailAdd{sameer.murthy@kcl.ac.uk}

\abstract{
We study invariants of bosonic and fermionic (Grassmann-valued) matrices under the adjoint 
action of~$U(N)$, weighted by the fermion number.
Such models naturally appear as the supersymmetric indices of supersymmetric gauge theories and are captured by $U(N)$ matrix models.
We discuss two features of the fermionic models that are qualitatively different from bosonic models. 
Firstly, the~$2N^\text{th}$ power of a Grassmann matrix vanishes, which gives rise to many 
new trace relations. Secondly, trace relations in models involving fermions could cause an 
increase in the supersymmetric index  
as~$N$ decreases, in contrast with purely bosonic models. 
We focus on a simple model involving one fermion and one derivative that corresponds 
to a~$\frac14$-BPS supersymmetric index in~$\CN=4$ SYM theory, in which 
we find that the index is independent of~$N$. 
We prove this rank-independence analytically, and 
experimentally study the cancellations between bosonic and fermionic trace relations that lead to it.
Based on these observations, we make some conjectures on resulting algebraic structures, 
including the analogue of the polarized Cayley-Hamilton identities and the Second Fundamental Theorem of invariants in the presence of Grassmann matrices. 
Finally, we present various (smooth and singular) limits of the most general 
supersymmetric index in~$\CN=4$ SYM theory, and study some patterns in their behavior as 
a function of~$N$.
}

\begin{document}

\maketitle 

\section{Introduction}

The enumeration and classification of invariant functions of a given set of~$N \times N$ complex-valued matrices  
under the adjoint action of~$U(N)$, or $GL(N)$,  is a well-studied problem that appears in many areas of 
physics and mathematics, see, e.g.~\cite{Morozov:2009jv,de2017invariant}. 
At~$N=\infty$, the ring of invariants is freely generated by the traces of all monomials in the 
matrices.
When~$N < \infty$, trace relations come into effect and introduce relations among these generators. 
In this paper we discuss a variant of this enumeration problem when some of the matrices are 
\emph{fermionic}.\footnote{Throughout this paper we refer to matrices whose elements are 
Grassmann numbers as fermionic matrices, and  
those whose elements are complex numbers as bosonic matrices.}

Our physical motivation for studying such enumeration problems comes from the fact that they can 
be regarded as a zero-dimensional version of gauge theories that model interesting features 
of their phase structure~\cite{Sundborg:1999ue,Polyakov:2001af,Aharony:2003sx}.  
More precisely, the partition functions of gauge theories on compact spaces can be reduced to 
integrals over the gauge group, which we take to be $U(N)$ in this paper. 
Given that fermionic matrices arise very naturally in the context of adjoint-valued matter in gauge theories,
including supersymmetric gauge theories, 
one is naturally led to the study of fermionic matrix invariants. 
Related aspects of fermionic matrix models in physics have been studied in~\cite{Makeenko:1993AdjointFermion,
Semenoff:1996FermionicMatrix,
Chryssomalakos:1998gh,
Paniak:2000zy, deAzcarraga:2000rj, 
Anninos:2015eji,Anninos:2016klf,
Klebanov:2018nfp,
Gaitan:2020zbm,
Troost:2020tdd,Chen:2025sum}, and mathematical studies in the context of fermionic Lie algebra and invariants of 
supermatrices\footnote{In this paper 
we consider matrix invariants under the adjoint action 
of~$U(N)$ (or, equivalently, of $GL(N)$). In particular, even though we consider fermionic matrices, 
we do not consider invariants under the adjoint action of supermatrices.} include \cite{Chevalley:1948zz,
Koszul:1950,
Kostant1965,
Berele:1988GradedInvariants,
itoh2015,
Berele:2013GrassmannAlgebra}.
Our main focus of study here is on two related features of fermionic matrix invariants, 
which are strikingly different from the analogous bosonic problem. 
Both these features have to do, primarily with relations between invariants. 

\medskip

In order to state the first feature, we first recall some well-known facts about bosonic 
matrices. 
In the simplest case when there is only one matrix~$X$, 
the Cayley-Hamilton theorem implies that~$\Tr \, X^{N+1}$ is a polynomial in~$\{\Tr \, X^n; n =1,\dots, N\}$. 
For multiple matrices, the problem is much more complicated. 
There is a large literature in algebra dedicated to the study of this problem, beginning with~\cite{Razmyslov:1974TraceIdentities,Procesi:1976Invariant}. 
It is known that the relations still take the form of the Cayley-Hamilton identity  
for multiple matrices, but now there can be relations-for-relations, etcetera.

When the elements of an~$N \times N$ matrix~$\Psi$ are Grassmann valued, 
the picture is somewhat different. 
The space of all functions of the matrix is finite-dimensional because of the nilpotency of the elements. 
One may therefore expect new relations compared to bosonic matrices. 
For example, since there are~$N^2$ independent elements in $\Psi$, it obeys~$\Psi^{N^2+1}=0$, 
and so any traces involving such large powers vanishes.
The surprise is that such a vanishing occurs at much smaller powers: 
as we discuss in Section~\ref{sec:singlefermion}, $\Psi^{2N}=0$.
In other words, while, for a matrix~$X$ whose elements are complex numbers, 
the powers~$X^{N+i}$, $i=1,2,\dots$ are algebraic functions of the lower powers, 
the~$2N^\text{th}$ power of a Grassmann-valued matrix simply vanishes. Inserting this relation inside traces leads to relations between invariants that have been studied previously. 

\medskip

The second feature is an application of fermionic trace relations to the~$N$-dependence 
of supersymmetric indices, as we explain presently after briefly setting up the context. 
Consider the quantum-mechanical model consisting of a set of~$N \times N$ matrix-valued 
harmonic oscillators of unit frequency, that are gauged with respect to the adjoint action of~$U(N)$,
and with no other interactions. 
The physical Hilbert space~$\mathcal{H}$ of the model is a subspace of the Fock space of multiple oscillators corresponding to the multiple matrices. 
The elements of~$\mathcal{H}$ are in one-to-one correspondence with gauge-invariant functions of 
the matrices, and has the structure of a ring graded by the eigenvalue of the Hamiltonian.

We study the number of gauge-invariants as encoded in traces of the following form,  
\be \label{eq:ZNIN}
Z_N (q) \= \Tr_{\mathcal{H}} \, q^L  \,,  \qquad 
I_N (q) \= \Tr_{\mathcal{H}} \, (-1)^F \,  q^L \,. 
\ee 
Here,~$L$ is the Hamiltonian shifted by an additive constant such that the~$L$-eigenvalue of the vacuum vanishes. 
The~$L$-eigenvalue of a state in the Fock space is thus simply the number of raising operators acting on the vacuum in that state. 
From the physical point of view, with~$q=e^{-\gamma}$, $Z_N$ is essentially the thermal partition function 
of this quantum-mechanical theory at inverse temperature~$\gamma$.\footnote{One can also 
turn on more general chemical potentials in the theory within the same formalism, but we do not study that here.} 
The trace $I_N(q)$ differs from the partition function only in that fermions pick up 
a negative sign. 
Note that this is not the supersymmetric index of the above model and, in general, 
there is no reason that the  quantum-mechanical model is even supersymmetric.\footnote{
It is, of course, possible that some of these models have their own supersymmetric structure, as is the case in the single fermion model studied below.}
But, as we explain presently, it can arise as the supersymmetric index of a certain supersymmetric gauge theory. 
Henceforth we shall refer to~$Z_N$ and~$I_N$ simply as the partition function and the index, respectively.

\smallskip

Our primary motivation and source of examples of such quantum-mechanical models 
comes from four-dimensional $U(N)$ gauge theory with adjoint-valued matter on~$S^3$. 
The free gauge theory problem reduces to a quantum-mechanical problem of the above sort,
where the (infinite) set of matrices are the Kaluza-Klein harmonics of all the fields 
on~$S^3$~\cite{Sundborg:1999ue,Aharony:2003sx}.
When the gauge theory is supersymmetric, $I_N(q)$ arises as the supersymmetric index of the 
gauge theory~\cite{Romelsberger:2005eg,Kinney:2005ej}, 
graded by a conserved charge~$L$ that commutes with the supercharge of the gauge theory. 
The partition function and the index can be expressed and calculated in different ways.
\begin{enumerate}
\item In the simplest cases, one may have at hand an explicit list of all the gauge-invariants 
of the quantum-mechanical model.
In such cases, one can directly calculate the traces in~\eqref{eq:ZNIN}, which gives valuable insight. 

More commonly, one does not have a simple way of explicitly listing the gauge-invariants,
but one has a knowledge of the fields entering the gauge theory. 
In such cases, one can still enumerate the gauge-invariants of the free theory using two different methods. 
\item 
A well-known method to enumerate invariants is in terms of a~$U(N)$ matrix model. 
One ascribes a \emph{single letter} to each field, derivative, and constraint in the original problem, and calculates 
the \emph{single-letter partition function}~$z(q)$ and 
the \emph{single-letter index}~$i(q)$, 
defined as the same traces as in~\eqref{eq:ZNIN} but over the space of single letters.
The index of the full theory is then given by the  Molien-Weyl formula \cite{Teranishi,Pouliot:1998tk,Kinney:2005ej,Benvenuti:2006qr} 
\begin{equation}\label{eq:intU}
       I_N(q) \= \int_{U(N)} dU \exp \Bigl( \, \sum_{k=1}^{\infty}\frac{1}{k} \, i(q^k)\, \Tr\, U^k\, \Tr \, U^{-k} \Bigr) \,,
\end{equation}
where~$dU$ is the invariant (Haar) measure of~$U(N)$.
\item 
Another useful method to calculate the partition function and index  
 involves a sum over partitions \cite{Dolan:2007rq,Murthy:2022ien}.  
We use the frequency notation for partitions
$\boldsymbol\lambda =1^{k_1} \,  2^{k_2}\ldots \,$, 
$|\boldsymbol\lambda|=\sum_j j k_j$ to denote the weight of the partition, 
and $\ell(\boldsymbol\lambda)=\sum_j k_j$ to denote the length (number of non-zero parts) of the partition.
Then the index can be written as 
\begin{equation}\label{eq:char_exp1}
    I_N(q) \= \sum_{ \boldsymbol\lambda}\frac{i_{\boldsymbol\lambda}(q)}{z_{\boldsymbol\lambda}} \;
    \sum_{\boldsymbol\mu \atop    \ell(\boldsymbol\mu)\leq N } \chi^{\boldsymbol\mu} ( \boldsymbol\lambda)^2 \,,
\end{equation}
where~$z_{\boldsymbol\lambda} =  \prod_{j\ge 1} k_j! \; j^{k_j}$ 
and~$i_{\boldsymbol\lambda}(q) = \prod_{j\geq 1} i(q^{j})^{k_j}$,
and~$\chi^{\boldsymbol\mu} ( \boldsymbol\lambda)$ are the characters of the symmetric group.
\end{enumerate}
The free partition function~$Z_N(q)$ is given by the formulas of the type~\eqref{eq:intU}, \eqref{eq:char_exp1} 
with~$i(q)$ replaced by~$z(q)$.\footnote{The index and the partition function depend 
on the theory through these single-particle indices only. We use different notations for this dependence,
and sometimes suppress it in the notation.}
 
We make use of all these three methods in the bulk of the paper.\footnote{See~\cite{Rossello:2026} 
for a nice summary of the three methods as applied to the~$\frac12$-BPS index.}
Our interest here is in the dependence of the index on the 
rank~$N$~\cite{Murthy:2020scj,Murthy:2022ien}.\footnote{The dependence of the BPS states
themselves on the rank~$N$ is also a very interesting problem involving chaotic 
behavior~\cite{Chang:2024zqi,Chen:2024oqv,Chen:2025sum} that we do not study here. See \cite{Chen:2025sum,Tierz:2026vrs} for a discussion on BPS states in fermionic matrix models.} 
To be concrete, let us discuss the microcanonical ensemble, 
i.e.~the number of states (or index) $d_N(\ell)$ for fixed eigenvalue~$\ell$ of~$H$.
This is simply the~$\ell^\text{th}$ coefficient in the expansion
\be
I_N(q) \= \sum_{\ell \in \IZ} d_N(\ell) \, q^\ell \,.
\ee
There is a similar expansion for the free partition function which we discuss for some 
examples below.
In both cases, as~$N \to \infty$, the numbers~$d_N(\ell)$ stabilize to a number~$d_\infty(\ell)$ 
independent of~$N$,\footnote{In fact, for a large class of unitary matrix models including~\eqref{eq:intU},  
this stabilization occurs sharply across the diagonal~$\ell =  \alpha N$, 
where~$\alpha$ is the order of the power series~$i(q)$~\cite{Murthy:2022ien}.}
which are freely generated by the single traces. 
As we decrease~$N$ from~$\infty$, trace relations kick in. 

If all the signs in the single-letter partition function are positive, 
then simple trace relations among the generators of the free ring 
will decrease (more precisely, not increase) the number of invariants~$d_N$ as~$N$ decreases. 
Although there are relations for relations etc, the fact of decrease remains true in the full theory, as can be deduced 
from the expression~\eqref{eq:char_exp1} upon noting that all terms in the sum are positive in this case.
Related methods have been used to study the 
finite-$N$ structure of the Hilbert space in~\cite{Ramgoolam:2016ciq,deMelloKoch:2025rkw,Caputa:2025ikn}.

When there are fermionic fields in the problem, there can be negative signs in the 
single-particle partition function 
and the single-particle index.\footnote{Note that negative signs can 
appear due to elementary fermions, constraints or equations of motion.}
The problem then becomes more interesting, because fermionic trace relations will 
decrease the effective number of fermions, and hence will contribute towards an \emph{increase} 
in the value of~$d_N(\ell)$. 
If there are more fermionic than bosonic trace relations, the net value of~$d_N(\ell)$ will increase.  
If~$d_\infty(\ell)$ is positive, then~$d_N(\ell)$ will also increase in
absolute value as~$N$ decreases. 
This type of dependence on~$N$ is the second aspect of fermionic models 
that we study in this paper.

\medskip

In order to study this phenomenon, we consider different supersymmetric indices in $\CN=4$ 
Super Yang-Mills (SYM) theory. The simplest index, and the one that enumerates the most 
(super)symmetric states, is the~$\frac12$-BPS index.
In this case~$z(q)=i(q)=q$, which corresponds to the fact that the only single letter in the problem 
is a bosonic scalar. The corresponding enumeration problem is the simplest one involving one
bosonic matrix referred to earlier, and the first trace relation is simply 
the Cayley-Hamilton identity for one matrix at~$L=N+1$. 
Thus, the~$\frac12$-BPS index does not exhibit the fermionic dependence that we are interested in here. 

At the other end, we have the~$\frac{1}{16}$-BPS index, which is the most general supersymmetric 
index, and captures the least supersymmetric states. 
The~$\frac{1}{16}$-BPS index contains examples of a growth in magnitude as~$N$ decreases 
with~$\ell$ fixed~\cite{Murthy:2020scj}. 
Studying patterns in these relations would teach us about the detailed nature of the microstates 
of~$\frac{1}{16}$-BPS black holes in this theory. 
We believe that  fermionic trace relations are the underlying cause of the black-hole like growth of 
the~$\frac{1}{16}$-BPS index, but we have not yet managed to tame this structure. 
We leave this exciting example for future study and focus on simpler examples in this paper.

A model relatively less-studied in the literature is a~\RI-BPS index, which  preserves~8 supercharges. 
The single letters of this model are simply a fermion~$\psi$ and one derivative, which we denote by~$\p$. 
The single-letter index of this model has (only) negative signs (see Equation~\eqref{eq:slfdpsi}), 
and is a simple example that exhibits the phenomenon of fermionic trace relations described above.
In this case, we find that this index is \emph{independent} of~$N$. 
We present the proof of rank-independence of this index in Section~\ref{sec:fermionplusderiv}, 
using the so-called $q$-Dyson theorem applied to the integral representation~\eqref{eq:intU} of the index. 
In Appendix~\ref{app:unique} we show that the single-letter trace of the~\RI-BPS index is the unique one   
that is rank-invariant, up to a rescaling of~$q$. We list consequences and identities related 
to the proof in Section~\ref{sec:fermionplusderiv} and in Appendix~\ref{app:identities}. 
In Appendix~\ref{app:DSSYK}, we comment on certain interesting relations of this index to 
observables in the double-scaled SYK  model~\cite{Berkooz:2018qkz,Berkooz:2018jqr,Gaiotto:2024kze}.

Expressing the $N$-independence of the index in the language of the approaches~1 and~3 above 
leads to new insights. 
For example, upon using the expression for the index as a sum over partitions, 
one obtains non-trivial identities for the characters of the symmetric group. We also 
explain this in Section~\ref{sec:fermionplusderiv}. 
In the language of the first approach, the $N$-independence of the index implies 
that the number of bosonic trace relations exactly equals the number of fermionic trace relations at every value of~$N$. 
We discuss this pairing of relations and related algebraic aspects, such as theorems on trace invariants 
and relations~\cite{Procesi:1976Invariant,Procesi:2007LieGroups,Razmyslov:1974TraceIdentities}, 
polarized versions of the Cayley-Hamilton identity in Section~\ref{sec:polarizedCH}.  
Much of our discussion of this aspect of the problem is experimental, and it would be very interesting to  
study this in more detail and apply the insights to different supersymmetric indices of
$\CN=4$ SYM as well as of other theories.

\smallskip

With a view towards studying the~$N$-dependence of different indices, we present, in Appendix~\ref{app:Neq4indices},  
a classification of all limits of the most general, i.e.~$\frac{1}{16}$-BPS, index of~$\CN=4$ SYM. 
This has been presented earlier in the PhD thesis~\cite{Eleftheriou:2025nkz}. 
The study of different limits of~$\CN=4$ SYM indices is in the same spirit as the 
study of different limits of the~$\CN=2$ index studied in \cite{Gadde:2011uv}. 
We also discuss some interesting patterns in some indices of~$\CN=4$ SYM in Section~\ref{sec:expindices}.
We post related data on GitHub, see Appendix~\ref{app:supp} for a link.

\medskip

Our observations in Sections~\ref{sec:fermionplusderiv} and~\ref{sec:polarizedCH} 
point to the existence of a supercharge of the~$\p \psi$ model 
that pairs up the bosonic and fermionic operators (and relations), i.e.~with vanishing index. Such a supercharge does not descend in an obvious manner from~$\CN=4$ SYM theory.  
The single-fermion model, studied in   Section~\ref{sec:singlefermion} is known to have a supercharge, which is the Kostant cubic operator. 
It is worth exploring whether a similar supersymmetric structure exists in the~$\p \psi$ model and in more general supersymmetric indices of~$\CN=4$ SYM. 
With such a structure in hand, we would be in a position to study the growth of trace relations itself as an index problem! 

\smallskip
Finally, the $N$-dependence of the partition function and the index relates to different aspects of holography \cite{Arai:2019xmp,Murthy:2020scj,Agarwal:2020zwm,Gaiotto:2021xce,Murthy:2022ien,Lee:2023iil,Lee:2024hef,deMelloKoch:2025LoopSpace,Lee:2025veh}.
In the holographic dual of a conformal gauge theory, $1/N$ corrections to the~$N \to \infty$ limit of the theory 
generically correspond to quantum-gravitational effects to the semi-classical theory. 
The strict~$N=\infty$ spectrum (i.e.~$N=\infty$ at finite energies), which is generated by single traces, 
is dual to the spectrum of gravitons in the dual AdS space. Since $1/N$ controls the interactions, it is 
interesting to ask whether and how gravitational structures like black holes appear as coherent states
of the~$N=\infty$ spectrum. Finite-$N$ effects are also crucial in the appearance of D-branes, as shown 
by the giant graviton expansion~\cite{Arai:2019xmp}, \cite{Gaiotto:2021xce}, \cite{Murthy:2022ien}. 
Our investigations in this paper show that fermionic matrices and their trace relations may have an 
important role to play in finite-$N$ holography.

\section{Fermionic trace relations: the one-fermion model}\label{sec:singlefermion}

In this section we study a simple toy model consisting of a single~$N \times N$ fermionic 
matrix~$\psi$ with charge $L=1$,  
on which the gauge group~$U(N)$ has the adjoint action~$\psi \to U \psi U^{-1}$. 
The single-letter expressions for this model are~$z(q) = q$ for the partition function
and \hbox{$i(q) = - q$} for the index.  

This toy model is instructive because we can solve the trace relations explicitly 
and derive the partition function and index from this point of view. 
This exercise illustrates how fermionic matrices generate trace relations.
In the process, we also derive a simple theorem about the nilpotency of fermionic matrices.
This is the first feature of fermionic matrices mentioned in the introduction.\footnote{After obtaining this 
result, we realized that we have rederived a known result in ring theory~\cite{amitsur1950minimal}. However, in spite of 
the multiple appearance of fermions in physics, we could not find this result in the physics literature, 
and so we include a simple discussion here.}

\subsection{A theorem regarding vanishing of powers of fermionic matrices}

For any matrix~$\psi$, the gauge-invariant functions of charge~$\ell \in \IN$ can be written in the following form, 
for some~$m \in \IN$, $k_j \in \IN$, $j=1,\dots m$, 
\begin{equation} \label{eq:ggeinvpsi}
\tr\bigl(\psi^{k_1} \bigr) \, \tr \bigl( \psi^{k_2} \bigr) \, \cdots \, \tr \bigl(\psi^{k_m} \bigr) \, , 
\qquad k_1 \geq k_2 \geq \cdots \geq k_m \, , \qquad 
\sum_{j=1}^m k_j\=\ell \,.
\end{equation}
We use the notation~$(\psi^k)_{ij}$, $i,j=1,\dots, N$ to denote the elements of powers of~$\psi$, $\tr$ to denote the trace, 
and the standard convention that repeated indices are summed from~1 to~$N$.
The Grassmann nature of~$\psi$ means that every element 
anticommutes with every other element and, in particular, each element is nilpotent.

\medskip

Even at $N=\infty$, the set of invariants is smaller than~\eqref{eq:ggeinvpsi} because of some simple identities obeyed by
traces of powers of fermionic matrices, which we call fermionic constraints. 
Firstly, for odd~$k$, $\bigl( \tr \, \psi^k \bigr)$ is a Grassmann number and hence its square vanishes. 
Secondly, the trace of any even power of~$\psi$ vanishes: 
\be
\tr \, \psi^{2k} \= \psi_{ij} \, \bigl( \psi^{2k-1} \bigr)_{ji} \=  \, \psi_{ji}  \bigl( \psi^{2k-1} \bigl)_{ij} 
\=  - \bigl( \psi^{2k-1} \bigl)_{ij} \, \psi_{ji}  \= -\tr \, \psi^{2k} \= 0 \,.
\ee
Here the first and fourth equalities hold by the definition of the trace, 
the second equality holds by relabelling indices~$i \leftrightarrow j$, and the third equality 
follows from the anticommuting nature of the elements of~$\psi$. 
We summarize the fermion constraints as follows, 
\be \label{eq:tracepsi}
\begin{split}
\tr \, \psi^k \= 0 \,, & \qquad \text{$k$ even} \,, \\ 
\bigl( \tr \, \psi^k \bigr)^2 \= 0 \,, & \qquad \text{$k \in \IN$} \,. \\
\end{split}
\ee
These constraints mean that the natural numbers~$k_j$ in~\eqref{eq:ggeinvpsi} are all distinct and odd, no matter what $N$ is. 

\medskip

For $N = \infty$, \eqref{eq:tracepsi} constitute the only relations. 
The number of such operators 
is therefore equal to the number of partitions of $\ell$ into odd distinct parts. 
The corresponding generating function is given by 
\begin{equation} \label{eq:Zinfonefer}
Z^\psi_\infty(q) \= \prod_{i=1}^{\infty} (1+q^{2i-1})\=(-q;q^2)_\infty \, ,
\end{equation}
where we have introduced the $q$-Pochhammer symbol $(x;q)_n = \prod_{i=0}^{n-1} (1-x \, q^i)$. 
To calculate the index, we simply notice that, since~$\psi$ has charge~1, 
the operator is bosonic (fermionic) when the number of parts is even (odd).
The generating function for the index is therefore given by 
\begin{equation} \label{eq:Iinfonefer}
I^\psi_\infty(q)\= \prod_{i=1}^{\infty} (1-q^{2i-1}) \= (q;q^2)_\infty \, .
\end{equation}

The result~\eqref{eq:Iinfonefer} can also be obtained from the character formula~\eqref{eq:Iinf1}, 
with $i(q)=-q$, as the following short calculation shows,  
\begin{equation}
\label{eq:checkIinf}
\begin{split}
    I^\psi_{\infty}(q)
    &\= \prod_{i=1}^{\infty}\frac{1}{(1+q^i)}
    \= \prod_{i=1}^{\infty} \frac{1}{(1+q^i)} \, 
    \frac{\prod_{i=1}^{\infty}(1-q^{2i-1}) \, (1-q^{2i})}{\prod_{i=1}^{\infty}(1-q^i)}\\
    &\= \prod_{i=1}^{\infty}\frac{(1-q^{2i-1}) \, (1-q^{2i})}{(1+q^i) \, (1-q^{i})}
    \=\prod_{i=1}^{\infty}(1-q^{2i-1}) \,.
\end{split}
\end{equation}
All infinite products above converge absolutely for $\abs{q}<1$.

\medskip

Now we look at finite~$N$. 
We begin by recalling the Cayley-Hamilton theorem. 
The characteristic polynomial $p_M(\lambda)$ of an $N \times N$ matrix $M$ is defined as 
\begin{equation}\label{eq:chpoly}
	p_M(\lambda) \= \det(\lambda-M)\= \sum_{r=0}^N (-1)^r \sigma_r(M) \,\lambda^{N-r} \,.
\end{equation}
The coefficients~$\sigma_r(M)$ are given by the elementary symmetric functions of the eigenvalues of~$M$,  
and can be written in terms of powers of the trace of~$M$ as follows, 
\begin{equation} \label{eq:defsigmar}
\sigma_r(M)\=\sum_{|\boldsymbol\mu|=r}\frac{(-1)^{r+\ell(\boldsymbol\mu)}}{z_{\boldsymbol\mu}}
\prod_{j\geq 1}\left(\tr(M^j)\right)^{k_j} \,,
\end{equation}
with the notation for partitions as given near~\eqref{eq:char_exp1}.
In particular, we have~$\sigma_0(M)=1$,  
$\sigma_1(M)=\tr(M)$, and $\sigma_N(M)=\det(M)$. 
Then the Cayley-Hamilton theorem states that $p_M(M)=0$.
In particular, this means that $N^\text{th}$ power of an~$N \times N$ matrix is expressible in terms of 
the lower powers, 
\be
M^N \= -\sum_{r=1}^{N} (-1)^{r} \sigma_r(M) \, M^{N-r} \,.
\ee

Now we apply this theorem to our model. A non-trivial fact, not very much used in physics, is that the Cayley-Hamilton 
theorem applies to matrices over any commutative ring. In particular, we can apply it to the matrix~$\psi^2$,
which has all mutually commuting entries. 
Now, the fact that traces of even powers of~$\psi$ vanish (the second line of~\eqref{eq:tracepsi})
implies that the functions~$\sigma_r(\psi^2)$ given in~\eqref{eq:defsigmar} identically vanish for~$r>0$! 
The Cayley-Hamilton theorem then reduces to 
\be \label{eq:CHfer}
\psi^{2N} \= 0 \,.
\ee 
As we discovered in the course of writing this paper, this theorem was discovered in 1950 by 
Amitsur-Levitzki~\cite{amitsur1950minimal}. 
The above proof is essentially the same one as given by Procesi more recently~\cite{procesi2013theorem}.

The expression~\eqref{eq:CHfer} is simple and powerful, and worth repeating in words: 
the powers of an arbitrary fermionic matrix vanish after order~$2N$,
which, of course, is much smaller than~$N^2$ as~$N \to \infty$.

\subsection{An explicit expression for the single-fermion index, and trace relations}

We can compare the number of trace relations generated by~\eqref{eq:CHfer} to that indicated 
by the finite $N$ index $I_N^\psi(q)$. 
In order to do that, we derive an explicit expression for $I_N^\psi(q)$.
We start from the general integral formula~\eqref{eq:intU}, which 
we write in terms of the eigenvalues~$z=(z_1,\dots,z_N)$ of~$U$ as follows,
\begin{equation} \label{eq:INintzi1}
    I_N(q) \= \frac{1}{N!}\int_{\mathbb{T}^N} \Delta_N(z) \, \Delta_N(z^{-1}) \,
    \exp \biggl( \,\sum_{k=1}^\infty \frac{f(q^k)}{k}\sum_{1\leq i, j \leq N}z_i^k \, z_j^{-k} \biggr) 
    \prod_{j=1}^N \frac{dz_j}{2\pi i \, z_j} \,,
\end{equation}
where~$\mathbb{T}^N=(S^1)^N$. Here we have introduced the Vandermonde determinants 
\begin{equation} \label{eq:defVand}
    \Delta_N(z) \= \prod_{1\leq i< j \leq N} (z_j-z_i) \,, \qquad \Delta_N(z^{-1}) 
    \= \prod_{1\leq i< j \leq N} (z_j^{-1}-z_i^{-1}) \,.
\end{equation}

For the single-fermion model, the general formula~\eqref{eq:INintzi1} with~$i(q)=-q$ leads to 
\begin{equation}
    I^{\psi}_N(q)\= \frac{1}{N!}\int_{\mathbb{T}^N} \Delta_N(z) \, \Delta_N(z^{-1}) \,
    \exp\biggl( \, \sum_{n=1}^\infty -\frac{q^n}{n}\sum_{1\leq i,j \leq N}z_i^n \, z_j^{-n} \biggr) \prod_{j=1}^N \frac{dz_j}{2\pi i \, z_j} .
\end{equation}
Upon resumming the exponent, and using the Taylor expansion of $\log(1-x)$, we obtain 
\begin{equation}
\begin{split}
     I^{\psi}_N(q)
     \=&\frac{1}{N!}\int_{\mathbb{T}^N} \Delta_N(z) \, \Delta_N(z^{-1}) 
     \prod_{1\leq i,\, j \leq N}(1-q\,z_i\,z_j^{-1})\,\prod_{j=1}^N \frac{dz_j}{2\pi i \, z_j} \= (q;q^2)_N\,.
\end{split}
\end{equation}
To prove the last equality, we first use the dual 
Cauchy identity~\cite{Macdonald1995SFHP} \footnote{Note a typo in eq $4.3'$ on page 65 in \cite{Macdonald1995SFHP}.}
to rewrite the integrand as
\begin{equation}
    \prod_{1\leq i,\, j \leq N}(1-q\,z_i\,z_j^{-1}) \=
     \sum_{\ell(\boldsymbol \lambda)\leq N}  (-1)^{|\boldsymbol\lambda|} \, q^{|\boldsymbol\lambda|} \, 
     s_{\boldsymbol \lambda}(z) \, s_{\boldsymbol\lambda^t}(z^{-1})\,,
\end{equation}
where $\boldsymbol \lambda$ is any partition and $|\boldsymbol \lambda|$ is the norm of partitions, 
$\ell$ is the length of the partitions; $\boldsymbol \lambda^t$ is the conjugate of the partition 
$\boldsymbol \lambda$, and $s_{\boldsymbol \lambda}(z)$ is the  Schur function. 
Then, using the orthogonality of Schur functions
\begin{equation}
    \frac{1}{N!}\int_{\mathbb{T}^N} \Delta_N(z) \, \Delta_N(z^{-1}) s_{\boldsymbol\lambda}(z)
    s_{\boldsymbol\mu}(z^{-1})\,\prod_{j=1}^N \frac{dz_j}{2\pi i \, z_j}\= \delta_{\boldsymbol\lambda,\,\boldsymbol\mu}\,,
\end{equation}
we arrive at a simple formula for the index
\begin{equation} \label{eq:Ipsiselfconj}
    I^{\psi}_N(q)
     \=\sum_{\substack{\text{self-conjugate }\boldsymbol \lambda \\ \ell(\boldsymbol \lambda)\leq N}}  
     (-1)^{|\boldsymbol\lambda|} \, q^{|\boldsymbol\lambda|} \,.
\end{equation}
From the Young tableaux, we see that the number of self-conjugate partitions of order $n$ and diagonal length $m$ 
is equal to the number of partitions of $n$ into $m$ \emph{distinct odd parts}. 
The generating for this quantity gives the final result for the index to be
\begin{equation} \label{eq:INonefer}
I^\psi_N(q)\= \prod_{i=1}^{N} (1-q^{2i-1}) \= (q;q^2)_N \, ,
\end{equation}
Similarly, for the finite-$N$ partition function, we get $Z^\psi_N(q) \=(-q;q^2)_N$. 

\medskip

We can also derive the same result from the representation as a sum over partitions~\eqref{eq:char_exp1}. 
First, recall how one derives the $\frac12$-BPS index with~$i(q)=q$:
\begin{equation}
\begin{split}
    I^{X}_N(q)
    & \= \sum_{ \boldsymbol\lambda} \, \frac{q^{| \boldsymbol\lambda |}}{z_{\boldsymbol\lambda}} \;
    \sum_{\ell(\boldsymbol\mu)\leq N } \chi^{\boldsymbol\mu} ( \boldsymbol\lambda)^2  
     \=   \sum_{| \boldsymbol\lambda |=1}^\infty 
    q^{| \boldsymbol\lambda |} \sum_{\ell(\boldsymbol\mu)\leq N } 
     \sum_{ \boldsymbol\lambda \vdash | \boldsymbol\lambda |} \, \frac{1}{z_{\boldsymbol\lambda}} \;
  \chi^{\boldsymbol\mu} ( \boldsymbol\lambda)^2 \\
    & \=   \sum_{| \boldsymbol\lambda |=1}^\infty 
    q^{| \boldsymbol\lambda |} \sum_{\ell(\boldsymbol\mu)\leq N }  \delta_{| \boldsymbol\lambda | , | \boldsymbol\mu|}  
     \=   \sum_{\boldsymbol\lambda  \atop \ell(\boldsymbol\lambda)\leq N }  
    q^{| \boldsymbol\lambda |} \=   \frac{1}{(q)_N} \,.
\end{split}
\end{equation}
To obtain the third equality we use the orthogonality relation of the characters of the symmetric group.  

Now, for the single fermion model, the formula~\eqref{eq:char_exp1} with~$i(q)=-q$ leads to 
\begin{equation}
\begin{split}
    I^{\psi}_N(q)
    & \=   \sum_{| \boldsymbol\lambda |=1}^\infty 
    q^{| \boldsymbol\lambda |} \sum_{\ell(\boldsymbol\mu)\leq N }  
    \sum_{ \boldsymbol\lambda \vdash | \boldsymbol\lambda |} \, (-1)^{\ell(\boldsymbol\lambda )} \, 
    \frac{1}{z_{\boldsymbol\lambda}} \;
  \chi^{\boldsymbol\mu} ( \boldsymbol\lambda)^2 \\
    & \=   \sum_{| \boldsymbol\lambda |=1}^\infty 
    (-q)^{| \boldsymbol\lambda |} \sum_{\ell(\boldsymbol\mu)\leq N }  
    \sum_{ \boldsymbol\lambda \vdash | \boldsymbol\lambda |} \, \sgn(\boldsymbol\lambda) \, 
    \frac{1}{z_{\boldsymbol\lambda}} \;
  \chi^{\boldsymbol\mu} ( \boldsymbol\lambda)^2 \,,
\end{split}
\end{equation}
as above, where we use~$ \text{sgn}(\boldsymbol\lambda) =(-1)^{|\boldsymbol\lambda|} \, (-1)^{\ell(\boldsymbol\lambda )}$ 
in the second step. 
We continue by expressing the right-hand side as  
\begin{equation}
\begin{split}
    I^{\psi}_N(q) & \=   \sum_{| \boldsymbol\lambda |=1}^\infty 
    (-q)^{| \boldsymbol\lambda |} \sum_{\ell(\boldsymbol\mu)\leq N }  
    \sum_{ \boldsymbol\lambda \vdash | \boldsymbol\lambda |} \, \frac{1}{z_{\boldsymbol\lambda}} \;
 \chi^{{\boldsymbol \mu}^t} ( \boldsymbol\lambda) \,  \chi^{\boldsymbol\mu} ( \boldsymbol\lambda)  \,,
\end{split}
\end{equation}
where $\mu^t$ is the conjugate (transpose) partition and $\chi^{\mu^t}$ is the character of
$S^\mu\otimes \sgn$, using the isomorphism $S^\mu\otimes \sgn \cong S^{\mu^t}$
(equivalently $\chi^{\mu^t}(\lambda)=\sgn(\lambda)\chi^\mu(\lambda)$) \cite{JamesKerberSymmetricGroup}.
Using the orthogonality relation of the characters of the symmetric group, 
we obtain, once again,~\eqref{eq:Ipsiselfconj}. 

\bigskip

Then, it is easy to see that the identity~\eqref{eq:CHfer}
implies that the independent single-trace operators at finite $N$ are
\begin{equation}\label{eq:onefsing}
    \tr\psi \,, \, \tr\psi^3 \,, \, \dots \,,\, \tr\psi^{2N-1} \,.
\end{equation}
If we assume that there are no other relations among \eqref{eq:onefsing}, or that the multitraces 
are freely generated by these single-trace operators \eqref{eq:onefsing}, 
then the finite $N$ partition functions and indices are
\begin{equation} \label{eq:ZNonefer}
Z^\psi_N(q) \= \prod_{i=1}^{N} (1+q^{2i-1})\=(-q;q^2)_N \, ,
\end{equation}
\begin{equation} \label{eq:INonefer2}
I^\psi_N(q)\= \prod_{i=1}^{N} (1-q^{2i-1}) \= (q;q^2)_N \, ,
\end{equation}
which matches the exact result. 
Note how similar the final step is compared to that of the matrix integral manipulations, 
as both result in freely counting \emph{distinct odd parts}. 
In one approach it follows from fermion statistics and trace relations, while in the other 
it follows from properties of self-conjugate partitions. 
This indicates that~\eqref{eq:CHfer} generates all the trace relations in the one-fermion model. 
This is consistent with the results on Lie-algebra valued fermions (see~\cite{Chryssomalakos:1998gh,deAzcarraga:2000rj,Troost:2020tdd} for a physics-oriented discussion and the relation to the Kostant cubic operator).
In particular, the highest power of fermions that is possible comes from having all the~$k_j=1$ in~\eqref{eq:ggeinvpsi}, and is~$1+3+\dots + 2N-1 = N^2$. 
This is nothing but the product of all the fermions~$\psi_{ij}$. 
We revisit these points in Section~\ref{sec:polarizedCH} from a more algebraic point of view.

\bigskip

We now unpack some consequences of these trace relations.
First of all, at rank~$N$, there are no non-trivial trace relations among different operators 
with charge smaller than~$2N+1$. 
This value of the charge should be contrasted with the corresponding value for the bosonic case, 
where the trace relations start to exist at charge $N+1$. 
This last fact is related to the statement that for charge~$n\le N$, the partition function of a 
bosonic matrix at rank~$N$ is the same as the partition function at~$N \to \infty$.

Further, the finite-$N$ canonical index $I_N^\psi(q)$ of the single-fermion model admits 
several elementary but useful properties. Firstly, it is a polynomial of degree $N^2$ with 
constant term~$+1$ and highest-degree term~$(-q)^{N^2}$.
The degree $N^2$ has a simple physical meaning: the matrix $\psi$ contains $N^2$ 
independent Grassmann-odd components, so the maximal possible charge is obtained 
by taking the product of all of them, and contributes $(-q)^{N^2}$ to the index.
Secondly, the coefficient of $q^\ell$ is non-negative for even $\ell$ and non-positive for odd $\ell$.
Indeed, in this model each letter $\psi$ carries both charge $L=1$ and fermion number $F=1$, 
so for any gauge-invariant operator we have $F\equiv L\pmod 2$, and therefore contributions 
at even (odd) charge are bosonic (fermionic) and enter the index with sign $+$ ($-$), respectively.
Finally, the sequence formed by the absolute values of the coefficients of any fixed degree as $N$ 
increases is non-decreasing. This is a direct consequence of the second property and the fact that 
the number of trace relations is non-increasing as~$N$ increases. 
(One can also prove these properties by induction in~$N$.)

\medskip

Upon expressing the $q$-Pochhammer symbol as
\begin{equation}
\label{eq:finitepochident}
    (x;q)_N \= \frac{(x;q)_\infty}{(xq^N;q)_\infty} \, ,
\end{equation}
and using the following identity~\cite{Zagier2007}, 
\begin{equation}
\label{eq:infinitepochident}
    \frac{1}{(x;q)_\infty}\=\sum_{m=0}^\infty \frac{x^m}{(q;q)_m} \,,
\end{equation}
we obtain 
\begin{equation} \label{eq:summary} 
    I^\psi_N(q)\=I^\psi_{\infty}(q) \, \sum_{m=0}^{\infty} \frac{q^m}{(q^2;q^2)_m}\, q^{2mN} \,.
\end{equation}
The equation~\eqref{eq:summary} can be regarded as the giant graviton expansion 
of this model.\footnote{The giant graviton expansion of the one-fermion model was also discussed in \cite{Lee:2023iil}.}

\section{Rank-invariance of \RI-BPS index}\label{sec:fermionplusderiv}

In this section we study a model that is a~\RI-BPS index of~$\CN=4$ SYM theory.  
It is defined as a certain limit of the general supersymmetric index in $\CN=4$ SYM
as we explain in Appendix~\ref{app:Neq4indices}.
The model illustrates the second feature of fermionic matrices mentioned in the introduction,
namely that while bosonic trace relations decrease the index as~$N$ decreases, 
fermionic trace relations work in the opposite manner. In fact, the index
is actually independent of~$N$, pointing to a curious balance of trace relations.
The model is simple to present. It consists of a single fermion~$\psi$ and 
a derivative~$\partial$, each carrying charge~$L=1$. We sometimes refer to this model as the~$\p \psi$ model.

\medskip

Before we consider this model, we briefly discuss some facts about the general matrix integral~\eqref{eq:intU}.
For $N=1$, the general matrix integral~\eqref{eq:intU} reduces to a trivial 
one-dimensional integral, which leads to  
\begin{equation} \label{eq:I11}
    I_1(q) \= \text{PE}(f(q))\, .
\end{equation}
It is a straightforward exercise to derive the same expression from the sum over partitions~\eqref{eq:char_exp1}.
At the other extreme, for $N=\infty$, the second sum in \eqref{eq:char_exp1} collapses, 
\begin{equation}
 \frac{1}{z_{\boldsymbol\lambda}}   \, \sum_{\boldsymbol\mu } \chi^{\boldsymbol\mu} ( \boldsymbol\lambda)^2 \= 1 \, ,
\end{equation}
and therefore we have
\begin{equation} \label{eq:Iinf1}
    I_{\infty}(q)\= \sum_{\boldsymbol\lambda} f_{\boldsymbol\lambda}(q)\= \sum_{\boldsymbol\lambda} \prod_{j\geq 1} f(q^{j})^{k_j} \= \prod_{j\geq 1} \sum_{n=0}^\infty f(q^{j})^{n}\= \prod_{j\geq 1} \frac{1}{1-f(q^j)} \, .
\end{equation}

\bigskip

Now we consider the index~$I^{\p \psi}_N(q)$ of the~$\p \psi$ model. 
The single-letter index is given 
by\footnote{Curiously, $i^{\p \psi}=-\widehat{f}=-\frac{f}{1-f}$ for~$f(q)=q$. 
Such an~$\widehat{f}$ is expected to appear as the single-particle index in the giant graviton expansion of~\cite{Murthy:2022ien} for the $\frac12$-BPS index of $\CN=4$ SYM theory.
This has been identified as the index of Koszul dual branes in the B-model topological string theory~\cite{Budzik:2025vzr}.  
Further, $\widehat{f}$ also encodes the fluctuations of the first giant correction of the~$\frac12$-BPS index~\cite{Gaiotto:2021xce}. This has been 
derived from the bulk D3-branes in AdS$_5$ in~\cite{Eleftheriou:2023jxr,Eleftheriou:2025lac}, where the ground state is fermionic and the 
states obtained by the action of the derivative are identified with the states of the lowest Landau level. 
We thank Kasia Budzik for interesting conversations on this topic. 
}
\begin{equation} \label{eq:slfdpsi}
    i^{\p \psi}(q) \= -\frac{q}{1-q} \= -q-q^2-q^3-q^4- \dots \, .
\end{equation}
From formula~\eqref{eq:I11} we have
\begin{equation}
    I^{\p \psi}_1(q)\= \text{PE} \bigl( i^{\p \psi}(q) \bigr)
    \=\exp \Bigl( \, \sum_{k=1}^\infty\frac{1}{k} i^{\p \psi}(q^k) \Bigr) 
    \= \prod_{n=1}^\infty(1-q^n) \= (q;q)_\infty    \,.
\end{equation}
Next we consider~$N=\infty$. 
From the formula~\eqref{eq:Iinf1} we have
\begin{equation}
    I^{\p \psi}_\infty(q)  \= \prod_{n=1}^\infty\frac{1}{1-i^{\p \psi}(q^n)} \= \prod_{n=1}^\infty (1-q^n) \= (q;q)_\infty \,,
\end{equation}
which coincides with the $N=1$ index.

The fact that~$I^{\p \psi}_1(q) = I^{\p \psi}_\infty(q)$ naturally leads to the question 
of what happens for intermediate values of~$N$. 
Initially, we found numerically up to large orders of charges that, in fact, the index does not depend on~$N$, i.e.,
\be \label{eq:RInv}
I^{\p \psi}_N(q) \= I^{\p \psi}_\infty(q) \,, \qquad N\=1,2,3,\dots \,. 
\ee
The case~$N=2$ can also be treated using some identities involving Pochhammer symbols,
using which we give a proof of~\eqref{eq:RInv} for~$N=2$ in Appendix~\ref{app:Neq2index}. 
In fact, with some effort, we also found a proof of this statement for all~$N$, using the 
matrix integral representation of the index, which we present below. 
The statement of rank-invariance of the index gives rise to interesting consequences. 
From the mathematical point of view, it leads to  identities among group characters, which we discuss after presenting the proof. 
From the physical point of view, it means that, at any $N$, the new trace relations appear 
in perfect bose-fermi pairs,  which we discuss in Section~\ref{sec:polarizedCH}.

\bigskip

Recalling that the~$N=\infty$ index is the multi-graviton index, and 
writing the multi-graviton index in terms of the single-graviton index as 
\be
 I^{\p \psi}_\infty(q) \= \text{PE} \bigl( i^{\p \psi}_\text{grav}(q) \bigr) \,,
\ee
we obtain the interesting relation 
\be
i^{\p \psi}_\text{grav}(q) \= i^{\p \psi}(q) \,.
\ee
The left-hand side of this equation is the index of single gravitons or, equivalently, single traces in the matrix model.
The right-hand side is the index of single letters in the matrix model.

\subsection{Proof of the rank invariance}

We now prove that $I^{\p \psi}_N(q) \=(q;q)_\infty$ for all $N$. The main idea is a series of matrix integral 
manipulations and the $q$-Dyson theorem, also called the Zeilberger–Bressoud 
theorem~\cite{zeilberger1985proof}, which originally evaluates the constant term of a $q$-analogue 
of Dyson's Laurent polynomial. Consider the integral expression for the index \eqref{eq:INintzi1} applied to the index  
defined by the single-letter index~\eqref{eq:slfdpsi}. 
Let us first write the~$(i,j)^\text{th}$ term in the exponent as 
\begin{equation}
\begin{split}
    \sum_{k=1}^\infty \, \frac{1}{k} \, i^{\p \psi}(q^k) \, (x_i/x_j)^k 
 &   \= - \sum_{k=1}^{\infty}  \, \frac{1}{k} \, \frac{q^k}{1-q^k} \, (x_i/x_j)^k   
  \= -\sum_{k=1}^\infty \sum_{n=1}^{\infty}  \, \frac{1}{k} \, q^{nk} \, (x_i/x_j)^k \\
&     \= \log \, \prod_{n=1}^\infty (1-q^n \, x_i/x_j) \,,
\end{split}
\end{equation}
where, to reach the second line, we have interchanged the order of summations. 
The integral expression~\eqref{eq:INintzi1} for the index thus takes the following form, 
\begin{equation} \label{eq:intIN1}
        I^{\p \psi}_N(q)\=\frac{1}{N!}\oint_{\mathbb{T}^N}\left(\prod_{i=1}^N \frac{dx_i}{2\pi i x_i}\right)
        \prod_{1\leq i<j\leq N}(x_i-x_j)(x_i^{-1}-x_j^{-1})  \prod_{i,j=1}^N \prod_{n=1}^\infty \bigl(1-q^nx_i/x_j \bigr) \,.
\end{equation}
After some minor manipulations, this can be expressed as follows, 
\begin{equation} \label{eq:intid1}
        I^{\p \psi}_N(q) \= \frac{(q;q)_\infty^N}{N!}\oint_{\mathbb{T}^N}\left(\prod_{i=1}^N \frac{dx_i}{2\pi i x_i}\right) \;  
        \prod_{i\neq j}^N(x_i/x_j;q)_\infty \,.
\end{equation}
The rank-invariance of the index is then equivalent to the following integral identity, for~$N = 1,2,\dots$, 
\begin{equation}\label{eq:intid}
\oint_{\mathbb{T}^N}\left(\prod_{i=1}^N \frac{dx_i}{2\pi i x_i}\right)\prod_{1\leq i< j\leq N}(x_i x_j^{-1};q)_\infty \, 
(x_j x_i^{-1};q)_\infty\=N! \,(q; q)_\infty^{1-N} \,,
\end{equation}
whose proof we turn to next.

\bigskip

The proof of~\eqref{eq:intid} takes the form of three equalities as follows\footnote{After writing this paper, 
Mart\'i Rossell\'o pointed out that essentially the same proof was given in~\cite{Bressoud}. 
We thank him for communicating it to us.}, 
\begin{equation}
\begin{aligned}\label{eq:3eqs}
& \oint_{\mathbb{T}^N}\left(\prod_{i=1}^N \frac{dx_i}{2\pi i x_i}\right) \, \prod_{1\leq i< j\leq N}(x_i x_j^{-1};q)_\infty(x_j x_i^{-1};q)_\infty\\
    & \quad \= \oint_{\mathbb{T}^N}\left(\prod_{i=1}^N \frac{dx_i}{2\pi i x_i}\right) \delta_N(x) \, \delta_N(1/x)
   \prod_{1\le i<j \le N} (q\, x_i x_j^{-1}; q)_\infty \, (q\, x_j x_i^{-1}; q)_\infty\\
   &  \quad \= N!\oint_{\mathbb{T}^N}\left(\prod_{i=1}^N \frac{dx_i}{2\pi i x_i}\right) \delta_N(x)\
   \prod_{1\le i<j \le N} (q\, x_i x_j^{-1}; q)_\infty \, (q\, x_j x_i^{-1}; q)_\infty\\
   & \quad  \= N! \, (q; q)_\infty^{1-N}\,.
\end{aligned}
\end{equation}
Here and below we use the notation 
$x = \{x_1,...,x_N\}$ and $x^{-1} = \{x_1^{-1},...,x_N^{-1}\}$, 
and the function 
$\delta_N(x)$ is defined as 
\be
\delta_N(x) \= \prod_{1\le i<j\le N} (1-x_i x_j^{-1} ) \,,  
\ee
and is sometimes called the normalized Weyl denominator.
Note that $\delta_N(x)$ is related to the Vandermonde determinant~\eqref{eq:defVand} by 
\begin{equation} \label{eq:DelVdMrel}
    \delta_N(x) \= \Delta_N(x) \, \prod_{j=1}^N x_j^{-(j-1)} \,.
\end{equation}

We now present the proof of the three equalities in~\eqref{eq:3eqs} in turn.
Firstly, we note the following identity that follows immediately from the definition of the $q$-Pochhammer symbol,
\begin{equation}\label{eq:qPoid}
    (a;q)_\infty\= (1-a) \, (q\, a;q)_\infty \,.
\end{equation}
Upon applying this identity to all the~$q$-Pochhammer symbols in the first line of~\eqref{eq:3eqs}, 
we obtain the first equality.

The third equality in~\eqref{eq:3eqs} results from the $q$-Dyson theorem, which states that~\cite{zeilberger1985proof} 
\begin{equation}
   \oint_{\mathbb{T}^N}\left(\prod_{i=1}^N \frac{dx_i}{2\pi i x_i}\right) 
   \prod_{1\le i<j\le N} (x_i x_j^{-1}; q)_\infty \, (q\, x_j x_i^{-1}; q)_\infty \= (q; q)_\infty^{1-N} \,.
\end{equation}
Using \eqref{eq:qPoid}, we can rewrite the $q$-Dyson theorem as 
\begin{equation}
   \oint_{\mathbb{T}^N}\left(\prod_{i=1}^N \frac{dx_i}{2\pi i x_i}\right) \prod_{1\le i<j\le N} (1-x_i x_j^{-1} ) 
   \prod_{1\le i<j \le N} (q x_i x_j^{-1}; q)_\infty \, (q\, x_j x_i^{-1}; q)_\infty \= (q; q)_\infty^{1-N} \,,
\end{equation}
which immediately leads to the last equality in \eqref{eq:3eqs}.

To obtain the second equality in~\eqref{eq:3eqs}, we prove the following integral identity for any permutation 
invariant function~$f(x_1,...,x_N)$, \footnote{If we regard $x_1,...,x_N$ as the eigenvalues 
of $U(N)$ matrices, then $f(x_1,...,x_N)$ is a class function of the $U(N)$ group.}
\begin{equation}\label{eq:lemma}
   \oint_{\mathbb{T}^N}\left(\prod_{i=1}^N \frac{dx_i}{2\pi i x_i}\right) \delta_N(x) \, f(x) \= 
    \oint_{\mathbb{T}^N}\left(\prod_{i=1}^N \frac{dx_i}{2\pi i x_i}\right) \frac{1}{N!} \,   \delta_N(x) \,  \delta_N(1/x) \,
    f(x)  \,.
\end{equation}
The proof of \eqref{eq:lemma} begins with a change of variables to write its left-hand side as
\begin{equation}
    \begin{split}
        I_f &\; \coloneqq \;\oint_{\mathbb{T}^N}\left(\prod_{i=1}^N \frac{dx_i}{2\pi i x_i}\right) \delta_N(x) \, f(x) 
        \= \oint_{\mathbb{T}^N}\left(\prod_{i=1}^N \frac{dx_i}{2\pi i x_i}\right) \delta_N(\sigma\cdot x) \, f(\sigma\cdot x)\\
       &\=\oint_{\mathbb{T}^N}\left(\prod_{i=1}^N \frac{dx_i}{2\pi i x_i}\right) \delta_N(\sigma\cdot x) \, f(x) \,,
    \end{split}
\end{equation}
where $\sigma\in S_N$ is any permutation of the $N$ variables. 
To obtain the last equality, we used the fact that $f(x)$ is permutation invariant $f(\sigma\cdot x) = f(x)$. 
Then we can take the average over all $N!$ permutations to get the same integral back:
\begin{equation}\label{eq:lemma2}
    I_f \=\frac{1}{N!}\oint_{\mathbb{T}^N}\left(\prod_{i=1}^N \frac{dx_i}{2\pi i x_i}\right) 
    \sum_{\sigma\in S_N}\delta_N(\sigma\cdot x) f(x) \,.
\end{equation}
The next step is to realize that 
\begin{equation}
    \sum_{\sigma\in S_N}\delta_N(\sigma\cdot x) \= \delta_N(x) \, \delta_N(1/x) \,.
\end{equation}
To see this, we use the relation~\eqref{eq:DelVdMrel} to the Vandermonde determinant, to write 
the sum of permutations as
\begin{equation}
    \sum_{\sigma\in S_N}\delta_N(\sigma\cdot x)
    \=\sum_{\sigma\in S_N}\Delta_N(\sigma\cdot x)\prod_{j=1}^N x_{\sigma(j)}^{-(j-1)}
    \=\Delta_N( x)\sum_{\sigma\in S_N}\text{sgn}(\sigma)\prod_{j=1}^N x_{\sigma(j)}^{-(j-1)} \,,
\end{equation}
where the last sum is nothing but the Vandermonde determinant for $1/x$:
\begin{equation}
    \sum_{\sigma\in S_N}\text{sgn}(\sigma)\prod_{j=1}^N x_{\sigma(j)}^{-(j-1)} \= \Delta_N(1/x) \,.
\end{equation}
Then, using $\delta_N(1/x)=\Delta_N(1/x)\prod_{j=1}^N x_{j}^{(j-1)}$,
we arrive at
\begin{equation}
    \sum_{\sigma\in S_N}\delta_N(\sigma\cdot x) \= \Delta_N(x) \, \Delta_N(1/x) \= \delta_N(x) \, \delta_N(1/x) \,.
\end{equation}
Plugging this identity into \eqref{eq:lemma2}, we obtain
\begin{equation}
    I_f\=\frac{1}{N!}\oint_{\mathbb{T}^N}\left(\prod_{i=1}^N \frac{dx_i}{2\pi i x_i}\right) \delta_N(x) \, \delta_N(1/x) f(x) \,,
\end{equation}
so that the identity \eqref{eq:lemma} is proved. 
Finally, choosing the permutation invariant function in~\eqref{eq:lemma} to be 
\begin{equation}
    f(x) \= \prod_{1\le i<j\le N} (q\, x_i x_j^{-1}; q)_\infty \, (q\, x_j x_i^{-1}; q)_\infty \,.
\end{equation}
we arrive at the second equality in~\eqref{eq:3eqs}.

\bigskip

\paragraph{Comments on uniqueness.} 
It is natural to ask whether 
 $I^{\p \psi}_N(q) $ is the unique unrefined $U(N)$ index that satisfies the condition of rank-invariance. 
The answer is affirmative. In fact, 
as we prove in 
Appendix~\ref{app:unique}, the condition~$I_1(q)=I_\infty(q)$ or, equivalently, $i_\text{grav}(q) = i(q)$, 
is enough to uniquely fix the index up to the rescaling~$q\to q^m$.

\paragraph{Comments on giant graviton expansion.} Note that the equation $I^{\p \psi}_N(q)=I^{\p \psi}_\infty(q)$ implies that the 
giant graviton expansion of the index for the $\partial \psi$ model is trivial, i.e. the branes do not give any net contribution. 
Since we have identified the index~$I^{\p \psi}$ as a certain $\frac{1}{4}$-BPS index of $\mathcal{N}=4$ SYM 
with gauge group $U(N)$, it would be interesting to find a bulk understanding of this trivial giant graviton expansion,
perhaps using similar techniques as those used in~\cite{Eleftheriou:2023jxr, Eleftheriou:2025lac}. 
Relatedly, since the index of the~$SU(N)$ theory is computed as the index of the~$U(N)$ theory divided by the index 
of the~$U(1)$ theory, the rank-invariance of the index for the~$\p \psi$ model shows that it has 
trivial~$SU(N)$ indices for all~$N$. 
It would be interesting to prove this directly using representation theory.\footnote{We thank Chris Beem for this suggestion.} 
Note, also, that having a trivial giant graviton expansion for all $N$ and having a trivial $SU(N)$ 
index for all~$N$ are equivalent.

\subsection{New character identities from rank-invariance }

The rank-invariance of the index expressed as a sum over partitions leads to interesting identities. 
The representation~\eqref{eq:char_exp1} as a sum over partitions leads to the following equation, for~$N=2,3,\dots$,  

\begin{equation}
\label{eq:INdiff}
   I_{N}(q) - I_{N-1}(q) \= \sum_{\boldsymbol\lambda}  f_{\boldsymbol\lambda} (q) \,\kappa_N(\boldsymbol\lambda)  \,,
\end{equation}
where
\begin{equation}
    \kappa_N(\boldsymbol\lambda) \=  \frac{1}{z_{\boldsymbol\lambda}} \sum_{\ell(\boldsymbol\mu)=N} \chi^{\boldsymbol\mu} (\boldsymbol\lambda)^2 \, .
\end{equation}
Since the weight of a partition is equal to at least its length, and since the character~$\chi^{\boldsymbol\mu} (\boldsymbol\lambda)$ 
is non-zero only for~$|\boldsymbol\mu|=|\boldsymbol\lambda|$, 
the nontrivial sum in \eqref{eq:INdiff} actually starts from $|\lambda|=N$. 
For~$i^{\p \psi}(q)=-q/(1-q)$, the difference between finite $N$ indices \eqref{eq:INdiff} can be written as
\begin{equation}\label{eq:Idiff}
    I^{\p \psi}_N(q)-I^{\p \psi}_{N-1}(q)\=i^{\p \psi}(q)\sum_{|{\boldsymbol\lambda}|\geq N} 
    \frac{1}{\Phi_{{\boldsymbol\lambda}}(q^{-1})}\kappa_N({\boldsymbol\lambda}) \,,
\end{equation}
where
\begin{equation}
    \Phi_{\boldsymbol\lambda}(q)\=\frac{(1-q)^{k_1}(1-q^2)^{k_2} \dots }{1-q}\qquad 
    ({\boldsymbol\lambda}=1^{k_1}2^{k_2} \dots ) \,.
\end{equation}

\bigskip

\ndt Then the rank-invariance of the index is equivalent to saying the difference between indices~\eqref{eq:Idiff} vanishes, 
which translates to the following compact algebraic identity,
\begin{equation} \label{eq:Phiresult}
    \boxed{\sum_{|{\boldsymbol\lambda}|\geq N} \frac{1}{\Phi_{{\boldsymbol\lambda}}(q^{-1})} \, 
    \kappa_N({\boldsymbol\lambda}) \= 0 \,.}
\end{equation}

Interestingly, the same function $\Phi_\lambda$ has appeared in the course of a study of finite group characters: 
it is simply the characteristic polynomial of~$\boldsymbol\lambda$ (thought of as a group element of the symmetric group~$S_n$, 
$n=|\boldsymbol\lambda|$) on the standard irreducible representation~{\bf St}$_n = \IC^n/\IC$ of~$S_n$~\cite{Zagier2004}.

In Tables~\ref{tab:N2} and~\ref{tab:N3} below, we list the first few terms for~$N=2,3$. In each case, for a given power of~$q$,
the sum in~\eqref{eq:Phiresult} terminates  after a finite number of terms and one can check  
the identity~\eqref{eq:Phiresult} explicitly.
\begin{table}[h]
    \centering
    \begin{tabular}{c|c|c|c|c}
        $|{\boldsymbol\lambda}|$ & ${\boldsymbol\lambda}$ & $\kappa_2({\boldsymbol\lambda})$ 
        & $\Phi_{\boldsymbol\lambda}(q)$ & $1/\Phi_{\boldsymbol\lambda}(1/q) + O(q^6) $ \\ \hline
        2 & $1^2$ & $1/2$ & $1-q$ & $-q - q^2 - q^3 - q^4 - q^5$\\
        & $2^1$ & $1/2$ & $1 + q$ & $q - q^2 + q^3 - q^4 + q^5$\\ \hline
        3 & $1^3$ & $2/3$ & $1 - 2 q + q^2$ & $q^2 + 2 q^3 + 3 q^4 + 4 q^5$ \\
        & $1^1 2^1$ & $0$ & $1 - q^2$ & $-q^2 - q^4$\\
        & $3^1$ & $1/3$ & $1 + q + q^2$ & $q^2 - q^3 + q^5$\\ \hline
        4 & $1^4$ & $13/24$ & $1 - 3 q + 3 q^2 - q^3$ & $-q^3 - 3 q^4 - 6 q^5$\\
        & $1^2 2^1$ & $1/4$ & $1 - q - q^2 + q^3$ & $q^3 + q^4 + 2 q^5$\\
        & $1^1 3^1$ & $1/3$ & $1 - q^3$ & $-q^3$\\
        & $2^2$ & $5/8$ & $1 + q - q^2 - q^3$ & $-q^3 + q^4 - 2 q^5$\\
        & $4^1$ & $1/4$ & $1 + q + q^2 + q^3$ & $q^3 - q^4$\\ \hline
    \end{tabular}
    \caption{$N=2$}
    \label{tab:N2}
\end{table}

\begin{table}[h]
    \centering
    \begin{tabular}{c|c|c|c|c}
        $|{\boldsymbol\lambda}|$ & ${\boldsymbol\lambda}$ & $\kappa_3({\boldsymbol\lambda})$ 
        & $\Phi_{\boldsymbol\lambda}(q)$ & $1/\Phi_{\boldsymbol\lambda}(1/q) + O(q^6) $ \\ \hline
        3 & $1^3$ & $1/6$ & $1 - 2 q + q^2$ & $q^2 + 2 q^3 + 3 q^4 + 4 q^5$ \\
        & $1^1 2^1$ & $1/2$ & $1 - q^2$ & $-q^2 - q^4$\\
        & $3^1$ & $1/3$ & $1 + q + q^2$ & $q^2 - q^3 + q^5$\\ \hline
        4 & $1^4$ & $3/8$ & $1 - 3 q + 3 q^2 - q^3$ & $-q^3 - 3 q^4 - 6 q^5$\\
        & $1^2 2^1$ & $1/4$ & $1 - q - q^2 + q^3$ & $q^3 + q^4 + 2 q^5$\\
        & $1^1 3^1$ & $0$ & $1 - q^3$ & $-q^3$\\
        & $2^2$ & $1/8$ & $1 + q - q^2 - q^3$ & $-q^3 + q^4 - 2 q^5$\\
        & $4^1$ & $1/4$ & $1 + q + q^2 + q^3$ & $q^3 - q^4$\\ \hline
        5 & $1^5$ & $61/120$ & $1 - 4 q + 6 q^2 - 4 q^3 + q^4$ & $q^4 + 4 q^5$\\
        & $1^3 2^1$ & $1/12$ & $1 - 2 q + 2 q^3 - q^4$ & $-q^4 - 2 q^5$\\
        & $1^2 3^1$ & $1/6$ & $1 - q - q^3 + q^4$ & $q^4 + q^5$\\
        & $1^1 2^2$ & $5/8$ & $1 - 2 q^2 + q^4$ & $q^4$\\
        & $1^1 4^1$ & $1/4$ & $1 - q^4$ & $-q^4$\\
        & $2^1 3^1$ & $1/6$ & $1 + q - q^3 - q^4$ & $-q^4 + q^5$\\
        & $5^1$ & $1/5$ & $1 + q + q^2 + q^3 + q^4$ & $q^4 - q^5$\\ \hline
    \end{tabular}
    \caption{$N=3$}
    \label{tab:N3}
\end{table}

We can also express the~$q$-coefficients of the  identity~\eqref{eq:INdiff} in terms of other special functions to obtain new identities. 
We illustrate this with some examples in Appendix~\ref{app:identities}. 

\bigskip

It is also interesting to present the $N$-invariance of the index as a character identity obtained from setting~$I_N(q)=(q)_{\infty}$.
From~$i^{\p \psi}(q) = -\dfrac{q}{1-q}$, and recalling that
$|\boldsymbol\lambda| = \sum_j j k_j$ 
and 
$\ell(\boldsymbol\lambda) = \sum_j k_j$, 
we obtain 
\be
f_{\boldsymbol\lambda}(q)
\= (-1)^{\ell(\boldsymbol\lambda)} \,
  \frac{q^{|\boldsymbol\lambda|}}{\prod_{i=1}^{\ell(\boldsymbol\lambda)} (1 - q^{\boldsymbol\lambda_i})}.
\ee
The statement 
$I_N(q)=I_\infty(q)=(q)_{\infty}$ for all $N \in \mathbb{N}$ can then be written as 
\be
\frac{1}{(q)_{\infty}}\sum_{\boldsymbol\lambda}  
  \frac{(-1)^{\ell(\boldsymbol\lambda)} \,q^{|\boldsymbol\lambda|}}{\prod_{i=1}^{\ell(\boldsymbol\lambda)} 
  (1 - q^{\boldsymbol\lambda_i})} \,\frac{1}{z_{\boldsymbol\lambda}} \sum_{\ell(\boldsymbol\mu)\leq N} 
  \chi^{\boldsymbol\mu} (\boldsymbol\lambda)^2 \= 1 \,, \qquad N = 1,2, 3,\dots \,.
\ee

\section{Polarized Cayley-Hamilton theorem and trace relations \label{sec:polarizedCH}}

We have seen in Section~\ref{sec:singlefermion} that all the trace relations at finite $N$ in the one-fermion model are generated by the fermion matrix nilpotency relation~\eqref{eq:CHfer}. 
To understand microscopically how the trace relations are generated and organized in the~\RI-BPS index 
(the~$\p \psi$ model) studied in Section~\ref{sec:fermionplusderiv} is more complicated, 
as the model contains infinitely many fermionic letters $\partial^n\psi$. 
In this section, we study the organizing principles for trace relations among  GL$_N$ invariants of bosonic and fermionic matrices.

When we have multiple  bosonic matrices over real or complex numbers, or more generally over any commutative ring, 
it is known that (i) the invariants, under the conjugation by the general linear group,  
of the functions of the matrices are generated by traces of monomials built out of the matrices; 
and (ii) the trace relations are generated by the  
so-called \emph{polarized Cayley-Hamilton theorem}.
These two statements are known as the First and the Second Fundamental Theorems, respectively,  in the invariant theory of matrices (see e.g.~the book \cite{de2017invariant}).

While the First and Second Fundamental Theorems of invariant theory are established in the 
bosonic setting and have been generalized in several super, graded settings~\cite{Regev:1987SignTrace,Berele:2007ColorTrace,Zhang:2020QuantumGLSupergroup,razzaq2025fundamental}, 
the theory of GL$_N$ invariants of purely Grassmann-valued matrices is less developed. 
The analogue of the First Fundamental Theorem 
for GL$_N$ invariants of purely Grassmann-valued matrices has been proven by Berele~\cite{Berele:1988GradedInvariants}. 
We could not find an established theorem regarding the analogue of the Second Fundamental Theorem in our precise setting.
Nevertheless, our discussion in Section~\ref{sec:singlefermion} suggests 
that the Second Fundamental Theorem also applies to the one-fermion model. 
Indeed, that discussion shows that the elements of the matrix $\psi^2$ are bilinears of Grassmann numbers, which are commutative, 
so the usual Cayley-Hamilton identity for $\psi^2$ at rank~$N$ leads to the fermion matrix nilpotency~\hbox{$\psi^{2N}=0$}. Further, our calculations of the number of invariants by independent methods agrees with 
the idea that this basic nilpotency generates all the trace relations in the one-fermion model. 

In fact, our computations in the $\p \psi$ model suggest that there is a more general story for the 
fermionic case that parallels the bosonic one, with crucial modifications due to permutation-dependent 
sign factors reflecting the anticommutativity of fermions. 
We will state our conjectures about invariants and trace relations of fermions in Section~\ref{sec:FTR}. 
A formal proof of these conjectures should be of independent mathematical 
interest.

\subsection{Polarized Cayley-Hamilton theorem and trace relations}

We begin by recalling the elements of the Cayley-Hamilton theorem for a single 
bosonic matrix. We use the conventions in~\cite{de2017invariant}.
Given an $N\times N$ matrix~$X$ over any commutative ring, 
we define the associated \emph{Cayley-Hamilton element of order $m=1,2,...$}, 
using the characteristic polynomial of a rank-$m$ matrix:
\begin{equation} \label{eq:CHelement}
    \text{CH}_m(X)\; \coloneqq \; X^m-\sigma_1(X)\, X^{m-1}+ \dots +(-1)^m\sigma_m(X) \,,
\end{equation}
where $\sigma_r(X)$ is the $r$-th elementary symmetric function computed using 
traces of powers of~$X$ as in \eqref{eq:defsigmar}. 
Note that the Cayley-Hamilton element is homogeneous of degree~$m$ because the 
elementary symmetric function~$\sigma_r(X)$ has degree~$r$. 
Then the classical Cayley-Hamilton theorem states that 
\be \label{eq:CHthm}
\text{CH}_N(X) \= 0 \,.
\ee
The Cayley-Hamilton elements satisfy the following recursion relation for $m=1,2,...$, 
\be \label{eq:CHrec}
\text{CH}_{m+1}(X) \= X\text{CH}_{m}(X)+(-1)^{m+1}\sigma_{m+1}(X)\,.
\ee
Since the elementary symmetric function $\sigma_r(X)$ vanishes when $r>N$,\footnote{The trace relations $\sigma_{r>N}(X)=0$ 
are elementary, and follows from the following simple identity, 
$$\det(I+tM)=\exp\big(\operatorname{tr}\log(I+tM)\big)=
\exp\!\left(\sum_{j\ge1}\frac{(-1)^{j-1}}{j}\,t^j\,\operatorname{tr}(M^j)\right)=\sum_{r\ge 0} \sigma_r(M)\,t^r\,.$$ 
The left-hand side is clearly a polynomial of degree~$N$ in~$t$, 
from which we conclude that $\sigma_{r>N}(X)=0$ and that the series on the right-hand side actually truncates.}  
the recursion~\eqref{eq:CHrec} shows that Cayley-Hamilton elements with order $m>N$ also vanish. 

\smallskip

The classical single-matrix trace relations are generated by the single-matrix Cayley-Hamilton 
theorem through the following form
\begin{equation}\label{eq:siglTrRel}
    \tr(X\,\text{CH}_N(X)) = 0\,,
\end{equation}
which expresses $\text{tr}\,(X^{N+1})$ as a combination of traces of lower powers of~$X$. 
This is a special case of the Second Fundamental Theorem of matrix invariants applied to a single matrix.

\bigskip

Now we turn to the \textit{polarized Cayley-Hamilton theorem}, which is the multi-matrix version 
of the classical Cayley-Hamilton theorem. We define an operation called polarization which maps 
degree-$m$ homogeneous functions of a single matrix to degree-$m$ homogeneous functions of $m$-matrices. 
It acts on elements as follows, 
\begin{equation}
\begin{aligned}\label{eq:polar_local}
& \text{Polarized}\left[\left(\tr (X)\right)^{k_1}\left(\tr (X^2)\right)^{k_2} \dots \left(\tr(X^h)\right)^{k_h} X^{m-r}\right]\\
        & \qquad \= \frac{1}{m!}\sum_{\pi\in S_m}\tr (X_{\pi(1)}) \dots \tr (X_{\pi(k_1)}) \times \tr (X_{\pi(k_1+1)}X_{\pi(k_1+2)}) \dots \\
        & \qquad\qquad \qquad\qquad \qquad \dots \tr(X_{\pi(r-h+1)} \dots X_{\pi(r)}) \times 
 X_{\pi(r+1)} \dots X_{\pi(m)} \,, \qquad 
\end{aligned}
\end{equation}
and is extended via linearity to linear combinations of the arguments,
so that the polarization of the~$r$-th term in the order-$m$ Cayley-Hamilton element~\eqref{eq:CHelement} is
\begin{equation} \label{eq:polar_sigma}
\begin{aligned}
    & \text{Polarized}\left[(-1)^r\sigma_r(X)X^{m-r}\right] \\
        & \quad \=  \sum_{|\bd\mu|=r}\frac{(-1)^{\ell(\bd\mu)}}{z_{\bd\mu}}\text{Polarized}
        \left[\left(\tr (X)\right)^{k_1}\left(\tr (X^2)\right)^{k_2} \dots \left(\tr(X^h)\right)^{k_h} X^{m-r}\right]\,.
\end{aligned}
\end{equation}
Here we use the standard notations for partitions as in the text below~\eqref{eq:char_exp1}.
One can see that the polarization eventually boils down to the sum over permutations of $m$ matrices 
with the trace brackets structure fixed. Then, applying the polarization procedure linearly to the Cayley-Hamilton element, 
we define the \emph{Polarized Cayley-Hamilton element of  order-$m$} in terms of~\eqref{eq:polar_sigma},
\eqref{eq:polar_local} as follows,
\be
\begin{split}\label{eq:pCH}
\text{PCH}_m(X_1,\dots,X_m)
        &\; \coloneqq \; \sum_{r=0}^m \text{Polarized}\left[(-1)^r\sigma_r(X)X^{m-r}\right] \,.
\end{split}
\ee
It is clear that the order-$m$ polarized Cayley-Hamilton element reduces to the 
usual single-matrix Cayley-Hamilton element when $X_1= \dots =X_m=X$.

The polarized Cayley-Hamilton element can also be written in one of the two following equivalent forms, 
thanks to the uniqueness of symmetric linear map with fixed diagonal, see e.g.~\cite{Procesi:1976Invariant}
\be
\begin{split}\label{eq:pCHeq}
\text{PCH}_m(X_1,\dots,X_m)
&\=\frac{1}{m!}\frac{\partial^m}{\partial t_1 \dots \partial t_m}\text{CH}_m(t_1 X_1+ \dots +t_m X_m)\Big|_{t_1= \dots =t_m=0}\\
        &\= \frac{1}{m!}\sum_{\epsilon_1,\dots,\epsilon_m\in\{0,1\}}(-1)^{m-\sum\epsilon_i} \, \text{CH}_m(\epsilon_1 X_1+\dots+\epsilon_m X_m)\,.
\end{split}
\ee
Using either of the above two expressions, the regular Cayley-Hamilton theorem~\eqref{eq:CHthm} 
leads to a multi-linear identity for $N\times N$ matrices, which is the polarized Cayley-Hamilton 
theorem for bosonic matrices $X_1,...,X_N$:
\begin{equation}
    \text{PCH}_N(X_1, \dots ,X_N)\=0\,.
\end{equation}

\bigskip

The Second Fundamental Theorem of invariants of multiple matrices says that the relations between invariants (i.e.~the trace relations)  
for any set of $N\times N$ matrices $\{X_1,...,X_{N+1}\}$ over commutative rings 
are all generated by the polarized Cayley-Hamilton identity through the following form:
\begin{equation}\label{eq:trPCH}
    \tr \bigl(X_{N+1} \, \text{PCH}_N(X_1, \dots ,X_N) \bigr) \= 0\,.
\end{equation}
This is the multi-matrix generalization of the classical single-matrix trace relation~\eqref{eq:siglTrRel} coming from the single-matrix Cayley-Hamilton theorem.

Note that, while the Second Fundamental Theorem establishes the complete set of trace relations, 
it does not provide a construction of the \emph{independent} trace relations. 
In fact, sometimes the theorem trivializes. 
For example, choosing $X_{N+1}$ to be the identity matrix results in no non-trivial trace relations, 
but rather tautological identities between trace functions. 
In general, the structure of the ring of invariants can be very complicated.  
This gap adds another difficulty in completely understanding the trace relations, on top of the 
already complex nature of the polarized Cayley-Hamilton 
theorem.\footnote{In fact, even the structure of trace relations for small number of letters is 
extremely non-trivial and there are only few exact results available in the math literature. We refer the readers to 
\cite{
Razmyslov:1974TraceIdentities,
Procesi:1976Invariant,
Teranishi,
nakamoto2002structure,
drensky2005generators,
aslaksen2006defining,
benanti2008defining,
drenskyLaScala2009lowdegree,
lopatin2012nilpotency,
hoge2012presentation,
derksen2017polynomial,
berele2022gl_n,
garciaMartinez2022ring,
eshmatov2025noncommutative}
for some of the historical and state-of-the-art results. 
Note that most of these classical results are for bosonic matrices.}

\subsection{Trace relations of fermions}\label{sec:FTR}

After reviewing the underlying theorems for bosonic trace relations, we now look at matrices 
over anti-commuting numbers. For a single $N\times N$ fermionic matrix $\psi$, 
we have seen in Section~\ref{sec:singlefermion} 
that the trace relations are generated by  $\tr(\psi^{2N+1})=0$, which is constructed from 
the somewhat surprising matrix identity $\psi^{2N}=0$. 
We have understood that this matrix identity is exactly the Cayley-Hamilton theorem for the matrix~$\psi^2$. 
In fact, there is another matrix identity of degree $2N-1$, which takes the form of the 
polarized Cayley-Hamilton theorem for one matrix $\psi$ and $N-1$ copies of $\psi^2$:
\begin{equation} \label{eq:PCHN}
    \text{PCH}_N(\psi,\psi^2, \dots ,\psi^2)\=0 \,.
\end{equation}
However, this matrix identity does not produce new independent trace relations in the single fermion model. 
Let us briefly illustrate this through some examples. 
The identity~\eqref{eq:PCHN} for~$N=2$ is
\begin{equation}\label{E:fer_mat_id1}
    \psi^3-\frac{1}{2}(\tr\psi) \, \psi^2-\frac{1}{2}(\tr\psi^3) \, \text{Id}_2\=0 \,.
\end{equation}
In order to try to produce new trace relations, we can take the trace of the left-hand side  multiplied by other arbitrary matrices. 
Firstly, it is easy to check that the trace of the left-hand side of~\eqref{E:fer_mat_id1} identically vanishes. 
Then, upon multiplying it with $\psi$ and taking the trace identically vanishes  
due to anti-commutativity, and, once again, this makes the trace relation trivial. 
Then, multiplying it with $\psi^2$ generates the same trace relation as multiplying $\text{CH}_2(\psi^2)$ by $\psi$. 
And so on... 
It is an amusing exercise to work out similar patterns for $N=3$, where~\eqref{eq:PCHN} takes the following form,  
\begin{equation}
    \psi^5-\frac{1}{3}(\tr\psi)\, \psi^4-\frac{1}{3}(\tr\psi^3)\, \psi^2-\frac{1}{3}(\tr\psi^5) \, \text{Id}_3 \= 0 \,.
\end{equation}
In Appendix \ref{app:supp}, we share a simple code to check fermionic trace and matrix identities.

The identities $\text{CH}_N(\psi^2)=\psi^{2N}=0$ and $\text{PCH}_N(\psi,\psi^2, \dots ,\psi^2)=0$ have been mentioned in~\cite{procesi2013theorem}, and have been called the basic even and basic odd identities in~\cite{brevsar2015quasi}.
In~\cite{brevsar2015quasi}, it has also been proved that there are no single fermion matrix identities of lower degree, 
which is consistent with our earlier discussion that trace relations in the one-fermion model are generated by $\text{CH}_N(\psi^2)=\psi^{2N}=0$.
We also need to comment that in the odd fundamental identity $\text{PCH}_N(\psi,\psi^2, \dots ,\psi^2)=0$, 
all the~$N$ matrices commute with each other and that is why the identity looks just like the bosonic identity. 
When one has both bosonic and fermionic  matrices 
it seems that the correct generalization of the polarized Cayley-Hamilton theorem should take 
the Grassmann nature into account in the form
of various signs. 
Through the study of the $\partial\psi$ model introduced in Section~\ref{sec:fermionplusderiv}, 
we observe patterns suggesting such a generalization. 
We summarize these patterns as two conjectures in dealing with fermionic trace relations.

The first is a Polarized Cayley-Hamilton conjecture for $N\times N$ matrices $M_1, \dots ,M_N$ 
where each $M_i$ can be either bosonic or fermionic.
Accordingly, we introduce the signed polarized CH elements for a combination of bosonic and fermionic matrices as follows, 
\begin{equation}\label{eq:generalizedCH1}
    \text{PCH}_N(M_1, \dots ,M_N) \; \coloneqq \; \sum_{i=0}^N \text{Polarized}_\text{sgn}\left[(-1)^i \sigma_i(M)M^{N-i}\right]  \,,
\end{equation}
where 
\begin{equation}
\begin{aligned}        &\text{Polarized}_\text{sgn}\left[(-1)^r\sigma_r(M)M^{m-r}\right]\\
        & \quad \coloneqq  \sum_{|\mu|=r}\frac{(-1)^{\ell(\mu)}}{z_\mu}\frac{1}{m!}\sum_{\pi\in S_m}\sgn(\pi;M_1, \dots ,M_m)\tr (M_{\pi(1)}) \dots \tr (M_{\pi(k_1)}) \dots\\
        &\qquad\qquad\qquad\qquad\qquad\qquad\qquad\qquad \tr(M_{\pi(r-h+1)} 
        \dots M_{\pi(r)})M_{\pi(r+1)} \dots M_{\pi(m)} \,.
\end{aligned}
\end{equation}
The sign function is defined by $\sgn(\pi;M_1, \dots ,M_m)=(-1)^{n_\pi}$ where $n_\pi$ 
is the number of swaps between fermionic matrices in order to implement the permutation $\pi$. 
\begin{conjecture}
With the above set-up, we have 
\begin{equation}\label{eq:generalizedCH}
    \text{PCH}_N(M_1, \dots ,M_N) 
    \= 0 \,.
\end{equation}
\end{conjecture}
\ndt The operation $\text{Polarized}_\text{sgn}$ is similar to the bosonic polarization~\eqref{eq:polar_local} except for 
the insertion of an additional sign function.
This is analogous to the so-called graded (Koszul) symmetrizer, and the sign function is called the Koszul sign, 
as we can assign degree~0 to bosons and degree~1 to fermions in this graded algebra. 
Note that when all $M_i$ are fermionic matrices, the sign function reduces to~$\sgn(\pi)$, 
i.e.~the sign function of the permutation. 
The generalized identity~\eqref{eq:generalizedCH} reduces to the polarized Cayley-Hamilton
theorem for the bosonic case when all the~$M_i$ are bosonic. 

\medskip

Our second conjecture is a version of the Second Fundamental Theorem for multiple matrices 
which can be bosonic or fermionic.
\vspace{-2mm}
\begin{conjecture}\label{conj:2}
    Relations among invariants (i.e.~the trace relations), are generated by the polarized 
    Cayley-Hamilton theorem of the more general form~\eqref{eq:generalizedCH}.
\end{conjecture}
\noindent We examined both these conjectures in the~$\p \psi$ model, and found the resulting 
observations consistent with them, as we proceed to discuss below.  
Along the way, we also make observations about patterns of trace relations in this model, 
which may hold independent interest.

\subsection{Patterns of trace relations in the $\partial\psi$ model}
\label{sec:trace-relations-dpsi}

In the $\partial\psi$ model introduced in Section~\ref{sec:fermionplusderiv}, 
we construct and organize trace relations in the form of the generalized polarized Cayley--Hamilton conjecture~\eqref{eq:generalizedCH}, 
and compare the resulting relations with the finite-$N$ counting of bosonic and fermionic operators obtained from matrix-integral formulas.
A striking empirical feature is a \emph{perfect pairing} between bosonic and fermionic trace relations at each rank and charge. 
In this section, we study numerical patterns in the $\partial\psi$ model with a focus on the pairing mechanism, which provides a microscopic perspective on the rank-invariance of the $U(N)$ index.

The explicit finite-rank bosonic/fermionic counts for this model for rank up to~5 
and charges up to~15
are summarized in Table~\ref{T:partition}. 
From these counts one can infer the number of independent trace relations at 
each~$(N,n)$, which we present in Table~\ref{T:relations}.
A notable feature of Table~\ref{T:relations} is the appearance of stabilized diagonals and a simple difference pattern that we observe  empirically:
\begin{itemize}
    \item \textbf{Pattern I (stabilized diagonals).} \label{patternI}
    For fixed $N$, the number of relations at charge~$3N$ and rank~$N$ equals the number of relations at charge~$3N+2k$  and rank~$N+k$, for all~$k\ge1$.

    \item \textbf{Pattern II (first difference on a neighboring diagonal).}\label{patternII}
    The difference between the number of relations at charge~$3N$ and rank~$N$ and the number of relations at charge~$3N-2$ and rank~$N-1$ equals~$4N-4$.
\end{itemize}
In the next few sections, we dive into Pattern I and II, list explicitly the trace relations generated by \eqref{eq:generalizedCH} at low charge and rank, compare them with the exact counting, observe and try to understand the pairing between bosonic and fermionic relations. 

We should say that we also find further patterns along the diagonals, for example, 
\begin{itemize}
    \item \textbf{Pattern III (second difference pattern).}\label{patternIII}
    The difference between the number of relations at charge $3N+1$ and rank $N$ and the number of relations at charge $3N-1$ and rank $N-1$ stabilizes to $32N-60$ for $N\ge3$.
\end{itemize}
However, we do not discuss Pattern III in this work. Interested readers can find the relevant data in appendix \ref{app:supp}. 
We believe that there are more patterns along the diagonals. 

\bigskip

Using computer algebra, we came up with two algorithms of listing trace relations: 
\begin{itemize}
    \item \emph{Bottom-up:} list all single-trace and multi-trace operators at fixed $(N,n)$, 
    solve for all linear dependencies by brute force, and extract an independent basis of relations.

    \item \emph{Top-down:} generate a spanning set of relations from the generalized polarized Cayley--Hamilton theorem,
    express them in the multi-trace basis, and row-reduce to isolate independent relations.
    A second reduction step identifies a minimal generating set in terms of basic PCH-generated identities.
\end{itemize}
In practice, we take the Top-down approach, so it is more straightforward to test our conjecture~\ref{conj:2}. It turns out both approaches reproduce exactly the bosonic/fermionic relation counts shown in Table~\ref{T:relations}. Our code used in this section is available in appendix \ref{app:supp}.

\begin{table}[h!]
\centering
\small
\begin{tabular}{ |p{1.3cm}||p{1.3cm}|p{1.3cm}|p{1.3cm}|p{1.3cm}|p{1.3cm}|p{1.3cm}| }
 \hline
 N & 1 & 2 & 3 & 4 & 5 & $\infty$\\
 \hline\hline
 $n=0$ & \blue{1, 0} & \blue{1, 0} & \blue{1, 0} & \blue{1, 0} & \blue{1, 0} & \blue{1, 0} \\
 \hline
 $n=1$ & \blue{0, 1} & \blue{0, 1} & \blue{0, 1} & \blue{0, 1} & \blue{0, 1} & \blue{0, 1} \\
 \hline
 $n=2$ & \blue{0, 1} & \blue{0, 1} & \blue{0, 1} & \blue{0, 1} & \blue{0, 1} & \blue{0, 1} \\
 \hline
 $n=3$ & 1, 1 & \blue{2, 2} & \blue{2, 2} & \blue{2, 2} & \blue{2, 2} & \blue{2, 2} \\
 \hline
 $n=4$ & 1, 1 & \blue{3, 3} & \blue{3, 3} & \blue{3, 3} & \blue{3, 3} & \blue{3, 3} \\
 \hline
 $n=5$ & 2, 1 & 6, 5 & \blue{7, 6} & \blue{7, 6} & \blue{7, 6} & \blue{7, 6} \\
 \hline
 $n=6$ & 2, 2 & 10, 10 & \blue{14, 14} & \blue{14, 14} & \blue{14, 14} & \blue{14, 14} \\
 \hline
 $n=7$ & 3, 2 & 16, 15 & 26, 25 & \blue{27, 26} & \blue{27, 26} & \blue{27, 26}\\
 \hline
 $n=8$ & 3, 3 & 25, 25 & 49, 49 & \blue{53, 53} & \blue{53, 53} & \blue{53, 53} \\
 \hline
 $n=9$ & 4, 4 & 40, 40 & 93, 93 & 107, 107 & \blue{108, 108} & \blue{108, 108}\\
 \hline
 $n=10$ & 5, 5 & 62, 62 & 169, 169 & 211, 211 & \blue{215, 215} & \blue{215, 215} \\
 \hline
 $n=11$ & 6, 6 & 94, 94 & 304, 304 & 413, 413 & 427, 427 & \blue{428, 428} \\
 \hline
 $n=12$ & 7, 8 & 142, 143 & 541, 542 & 808, 809 & 856, 857 & \blue{860, 861} \\
 \hline
 $n=13$ & 9, 9 & 212, 212 & 943, 943 & 1559, 1559 & 1702, 1702 & \blue{1717, 1717}\\
 \hline
 $n=14$ & 11, 11 & 312, 312 & 1625, 1625 & 2982, 2982 & 3381, 3381 & \blue{3433, 3433}\\
 \hline
 $n=15$ & 13, 14 & 456, 457 & 2769, 2770 & 5658, 5659 & 6710, 6711 & \blue{6876, 6877}\\
 \hline
\end{tabular}
\caption{Number of Bosonic, Fermionic operators at rank $N=1, \dots, 5$ and charge~$n$.}
\label{T:partition}
\end{table}
\begin{table}[h!]
\centering
\small
\begin{tabular}{ |p{1.3cm}||p{1.3cm}|p{1.3cm}|p{1.3cm}|p{1.3cm}|p{1.3cm}| }
 \hline
 N & 1 & 2 & 3 & 4 & 5 \\
 \hline\hline
 $n=0$ & 0, 0 & 0, 0 & 0, 0 & 0, 0 & 0, 0 \\
\hline
 $n=1$ & 0, 0 & 0, 0 & 0, 0 & 0, 0 & 0, 0 \\
\hline
 $n=2$ & 0, 0 & 0, 0 & 0, 0 & 0, 0 & 0, 0 \\
\hline
 $n=3$ & \textcolor{blue}{1, 1} & 0, 0 & 0, 0 & 0, 0 & 0, 0 \\
\hline
 $n=4$ & \textcolor{orange}{2, 2} & 0, 0 & 0, 0 & 0, 0 & 0, 0 \\
\hline
 $n=5$ & 5, 5 & \textcolor{blue}{1, 1} & 0, 0 & 0, 0 & 0, 0 \\
\hline
 $n=6$ & 12, 12 & \textcolor{blue}{4, 4} & 0, 0 & 0, 0 & 0, 0 \\
\hline
 $n=7$ & 24, 24 & \textcolor{orange}{11, 11} & \textcolor{blue}{1, 1} & 0, 0 & 0, 0 \\
\hline
 $n=8$ & 50, 50 & \textcolor{red}{28, 28} & \textcolor{blue}{4, 4} & 0, 0 & 0, 0 \\
\hline
 $n=9$ & 104,104 & 68, 68 & \textcolor{blue}{15, 15} & \textcolor{blue}{1, 1} & 0, 0 \\
\hline
 $n=10$ & 210, 210 & 153, 153 & \textcolor{orange}{46, 46} & \textcolor{blue}{4, 4} & 0, 0 \\
\hline
 $n=11$ & 422, 422 & 334, 334 & \textcolor{red}{124, 124} & \textcolor{blue}{15, 15} & \textcolor{blue}{1, 1} \\
\hline
 $n=12$ & 853, 853 & 718, 718 & \textcolor{brown}{319, 319} & \textcolor{blue}{52, 52} & \textcolor{blue}{4, 4} \\
\hline
 $n=13$ & 1708, 1708 & 1505, 1505 & 774, 774 & \textcolor{orange}{158, 158} & \textcolor{blue}{15, 15} \\
\hline
 $n=14$ & 3422, 3422 & 3121, 3121 & 1808, 1808 & \textcolor{red}{451, 451} & \textcolor{blue}{52, 52} \\
\hline
$n=15$ & 6863, 6863 & 6420, 6420 & 4107, 4107 & \textcolor{brown}{1218, 1218} & \textcolor{blue}{166, 166} \\
\hline
 $n=16$ & 13726, 13726 & 13084, 13084 & 9092, 9092 & \textcolor{yellow}{3130, 3130} & \textcolor{orange}{501, 501} \\
\hline
 $n=17$ & 27462, 27462 & 26539, 26539 & 19750, 19750 & 7761, 7761 & \textcolor{red}{1442, 1442} \\
\hline
 $n=18$ & 54959, 54959 & 53642, 53642 & 42254, 42254 & 18670, 18670 & \textcolor{brown}{3988, 3988} \\
\hline
 $n=19$ & 109918, 109918 & 108057, 108057 & 89214, 89214 & 43730, 43730 & \textcolor{yellow}{10625, 10625} \\
\hline
\end{tabular}
\caption{Number of Bosonic, Fermionic relations at ranks $N=1, \dots, 5$ and charges $n$. The stabilized diagonals with \hyperref[patternI]{Pattern I} are highlighted in blue. The diagonals satisfying \hyperref[patternII]{\textcolor{orange}{Pattern II}} are highlighted in orange. The diagonals satisfying \hyperref[patternIII]{\textcolor{red}{Pattern III}} are highlighted in red. We highlight other patterns in brown and yellow.}
\label{T:relations}
\end{table}

\subsubsection{Pattern I}

\paragraph{The first stabilized diagonal.} We now construct trace relations explicitly, starting from small ranks.
At rank $N=1$, the first nontrivial relations occur at charge $n=3$:
one fermionic relation and one bosonic relation:
\begin{equation}
\begin{aligned}
\tr\psi^3&\=0\\
\tr(\psi\,\partial\psi)-\tr(\psi)\,\tr(\partial\psi)&\=0\,.
\end{aligned}
\end{equation}
At rank $N=2$, the lowest relations again come as one fermionic and one bosonic relation at charge $n=5$:
\begin{equation}
\begin{aligned}
\tr\psi^5&\=0\,,\\
\tr(\psi^3\partial\psi)-\frac12\,\tr(\psi)\,\tr(\psi^2\partial\psi)-\frac12\,\tr(\psi^3)\,\tr(\partial\psi)&\=0\,.
\end{aligned}
\end{equation}
These are generated by the basic even and basic odd polarized Cayley--Hamilton 
identities~\cite{brevsar2015quasi} for a single fermion matrix $\psi$ for $N=1,2$ respectively.
Similarly, we find, in our data, a pairing of bosonic and fermionic trace relations along 
the \emph{first stabilized diagonal} in Pattern I at charge $n=2N+1$.
In each case, this pairing is generated by the pairing of 
the basic even and odd identities, which are 
\begin{equation}
\label{eq:basicDiagonalPairing}
\begin{aligned}
\tr\bigl(\psi\,\PCH_N(\psi^2,\psi^2,\dots,\psi^2)\bigr)&\=0\,,\\
\tr\bigl(\partial\psi\,\PCH_N(\psi,\psi^2,\dots,\psi^2)\bigr)&\=0\,.
\end{aligned}
\end{equation}
In this sense, the basic even and odd identities naturally explain the boson-fermion pairing of constraints.

\paragraph{Beyond the first stabilized diagonal.}
Now we look at the second stabilized diagonal in Pattern I. 
The starting point is charge $n=6$ at rank $N=2$, where we find $4$ fermionic and $4$ bosonic relations:
\begin{enumerate}
    \item \textbf{Fermionic:} 
    $\tr\psi\,\PCH_{\partial\psi,\partial^2\psi}$, 
    $\tr\psi\,\PCH_{\partial\psi,\psi^3}$, 
    $\tr\psi\,\PCH_{\psi^2,\psi\partial\psi}$, 
    $\tr\psi\,\PCH_{\psi^2,\partial\psi\psi}$.
    \item \textbf{Bosonic:} 
    $\tr\psi\,\PCH_{\partial\psi,\partial\psi\psi}$, 
    $\tr\psi\,\PCH_{\partial\psi,\psi\partial\psi}$, 
    $\tr\psi\,\PCH_{\psi^2,\partial^2\psi}$, 
    $\tr\psi\,\PCH_{\psi^2,\psi^3}$.
\end{enumerate}
The next entry on the same stabilized diagonal is charge $n=8$ at rank $N=3$, 
which again contains $4$ fermionic and $4$ bosonic relations:
\begin{enumerate}
    \item \textbf{Fermionic:} 
    $\tr\psi\,\PCH_{\psi^2,\partial\psi,\partial^2\psi}$, 
    $\tr\psi\,\PCH_{\psi^2,\partial\psi,\psi^3}$, 
    $\tr\psi\,\PCH_{\psi^2,\psi^2,\psi\partial\psi}$, 
    $\tr\psi\,\PCH_{\psi^2,\psi^2,\partial\psi\psi}$.
    \item \textbf{Bosonic:} 
    $\tr\psi\,\PCH_{\psi^2,\partial\psi,\partial\psi\psi}$, 
    $\tr\psi\,\PCH_{\psi^2,\partial\psi,\psi\partial\psi}$, 
    $\tr\psi\,\PCH_{\psi^2,\psi^2,\partial^2\psi}$, 
    $\tr\psi\,\PCH_{\psi^2,\psi^2,\psi^3}$.
\end{enumerate}
These examples display an explicit boson--fermion pairing at fixed charge: 
since $\psi$ and $\partial$ carry the same charge in this sector, 
one can often trade $\partial\psi$ insertions with $\psi^2$ insertions while preserving total 
charge but flipping statistics. 
This suggests a heuristic $\partial\leftrightarrow\psi$ exchange symmetry acting on the space 
of trace relations.
While we do not have a microscopic proof of such a symmetry, the rank-invariance of the index 
indicates that some exact pairing mechanism must be at work.

\paragraph{A map between diagonals via insertion of $\psi^2$.}
A second recurring structure is a map from relations at $(N,n)$ to relations at $(N+1,n+2)$ 
induced by inserting $\psi^2$ into the Cayley--Hamilton data:
\begin{equation}
\PCH_{A_1,A_2,\dots,A_m}\quad \longmapsto\quad \PCH_{A_1,A_2,\dots,A_m,\psi^2}\,.
\end{equation}
Under this map, the charge-$8$ rank-$3$ relations above are in one-to-one correspondence with 
the charge-$6$ rank-$2$ relations, 
and more generally this matches Pattern~I: each application increases the rank by $1$ and the charge by $2$. 
Motivated by this, we conjecture that the stabilized $(4,4)$ diagonal (equivalently, 
charge $n=2N+2$ at rank $N$) takes the universal form
\begin{enumerate}
    \item \textbf{Fermionic:} 
    $\tr\psi\,\PCH_{\psi^2,\dots,\psi^2,\partial\psi,\partial^2\psi}$, 
    $\tr\psi\,\PCH_{\psi^2,\dots,\psi^2,\partial\psi,\psi^3}$, 
    $\tr\psi\,\PCH_{\psi^2,\dots,\psi^2,\psi^2,\psi\partial\psi}$, \\
    $\tr\psi\,\PCH_{\psi^2,\dots,\psi^2,\psi^2,\partial\psi\psi}$.
    \item \textbf{Bosonic:} 
    $\tr\psi\,\PCH_{\psi^2,\dots,\psi^2,\partial\psi,\partial\psi\psi}$, 
    $\tr\psi\,\PCH_{\psi^2,\dots,\psi^2,\partial\psi,\psi\partial\psi}$, 
    $\tr\psi\,\PCH_{\psi^2,\dots,\psi^2,\psi^2,\partial^2\psi}$, \\
    $\tr\psi\,\PCH_{\psi^2,\dots,\psi^2,\psi^2,\psi^3}$.
\end{enumerate}

\bigskip

\bigskip

\subsubsection{Pattern II}
\paragraph{A neighboring diagonal and Pattern II.}
We now consider charge $n=7$ at rank $N=2$, where there are $11$ fermionic and $11$ bosonic relations:
\begin{enumerate}
\item \textbf{Fermionic:} 
$\tr\psi\,\PCH_{\partial\psi,\partial^3\psi}$,
$\tr\psi\,\PCH_{\partial\psi,\partial\psi\psi^2}$,
$\tr\psi\,\PCH_{\partial\psi,\psi\partial\psi\psi}$,
$\tr\psi\,\PCH_{\partial\psi,\psi^2\partial\psi}$,
$\tr\psi\,\PCH_{\psi^2,\partial^2\psi\psi}$,
$\tr\psi\,\PCH_{\psi^2,\psi\partial^2\psi}$,
$\tr\psi\,\PCH_{\psi^2,(\partial\psi)^2}$,
$\tr\psi\,\PCH_{\psi^2,\psi^4}$,
$\tr\psi\,\PCH_{\partial^2\psi,\psi^3}$,
$\tr\psi\,\PCH_{\partial\psi\psi,\partial\psi\psi}$,
$\tr\psi\,\PCH_{\partial\psi\psi,\psi\partial\psi}$.
\item \textbf{Bosonic:} 
$\tr\psi\,\PCH_{\partial\psi,\partial^2\psi\psi}$,
$\tr\psi\,\PCH_{\partial\psi,\psi\partial^2\psi}$,
$\tr\psi\,\PCH_{\partial\psi,(\partial\psi)^2}$,
$\tr\psi\,\PCH_{\partial\psi,\psi^4}$,
$\tr\psi\,\PCH_{\psi^2,\partial^3\psi}$,
$\tr\psi\,\PCH_{\psi^2,\partial\psi\psi^2}$,
$\tr\psi\,\PCH_{\psi^2,\psi\partial\psi\psi}$,
$\tr\psi\,\PCH_{\partial^2\psi,\partial\psi\psi}$,
$\tr\psi\,\PCH_{\partial^2\psi,\psi\partial\psi}$,
$\tr\psi\,\PCH_{\partial\psi\psi,\psi^3}$,
$\tr\psi\,\PCH_{\psi\partial\psi,\psi^3}$.
\end{enumerate}
The neighboring entry relevant for Pattern~II is charge $n=9$ at rank $N=3$, 
which contains $15$ fermionic and $15$ bosonic relations:
\begin{enumerate}
\item \textbf{Fermionic:} 
$\tr\psi\,\PCH_{\partial\psi,\psi^2,\partial^3\psi}$,
$\tr\psi\,\PCH_{\partial\psi,\psi^2,\partial\psi\psi^2}$,
$\tr\psi\,\PCH_{\partial\psi,\psi^2,\psi\partial\psi\psi}$,
$\tr\psi\,\PCH_{\partial\psi,\psi^2,\psi^2\partial\psi}$,
$\tr\psi\,\PCH_{\psi^2,\psi^2,\partial^2\psi\psi}$,
$\tr\psi\,\PCH_{\psi^2,\psi^2,\psi\partial^2\psi}$,
$\tr\psi\,\PCH_{\psi^2,\psi^2,(\partial\psi)^2}$,
$\tr\psi\,\PCH_{\psi^2,\psi^2,\psi^4}$,
$\tr\psi\,\PCH_{\psi^2,\partial^2\psi,\psi^3}$,
$\tr\psi\,\PCH_{\psi^2,\partial\psi\psi,\partial\psi\psi}$,
$\tr\psi\,\PCH_{\psi^2,\partial\psi\psi,\psi\partial\psi}$,
$\tr\psi\,\PCH_{\partial\psi,\partial^2\psi,\partial\psi\psi}$,
$\tr\psi\,\PCH_{\partial\psi,\partial^2\psi,\psi\partial\psi}$,
$\tr\psi\,\PCH_{\partial\psi,\partial\psi\psi,\psi^3}$,
$\tr\psi\,\PCH_{\partial\psi,\psi\partial\psi,\psi^3}$.
\item \textbf{Bosonic:} 
$\tr\psi\,\PCH_{\partial\psi,\psi^2,\partial^2\psi\psi}$,
$\tr\psi\,\PCH_{\partial\psi,\psi^2,\psi\partial^2\psi}$,
$\tr\psi\,\PCH_{\partial\psi,\psi^2,(\partial\psi)^2}$,
$\tr\psi\,\PCH_{\partial\psi,\psi^2,\psi^4}$,
$\tr\psi\,\PCH_{\psi^2,\psi^2,\partial^3\psi}$,
$\tr\psi\,\PCH_{\psi^2,\psi^2,\partial\psi\psi^2}$,
$\tr\psi\,\PCH_{\psi^2,\psi^2,\psi\partial\psi\psi}$,
$\tr\psi\,\PCH_{\psi^2,\partial^2\psi,\partial\psi\psi}$,
$\tr\psi\,\PCH_{\psi^2,\partial^2\psi,\psi\partial\psi}$,
$\tr\psi\,\PCH_{\psi^2,\partial\psi\psi,\psi^3}$,
$\tr\psi\,\PCH_{\psi^2,\psi\partial\psi,\psi^3}$,
$\tr\psi\,\PCH_{\partial\psi,\partial^2\psi,\psi^3}$,
$\tr\psi\,\PCH_{\partial\psi,\partial\psi\psi,\partial\psi\psi}$,
$\tr\psi\,\PCH_{\partial\psi,\partial\psi\psi,\psi\partial\psi}$,
$\tr\psi\,\PCH_{\partial\psi,\psi\partial\psi,\psi\partial\psi}$.
\end{enumerate}
We therefore have
\begin{equation}
(15+15)-(11+11)=4+4\,,
\end{equation}
in agreement with Pattern~II.

\paragraph{Two insertion maps: $\psi^2$ versus $\partial\psi$.}
The increase from $(N,n)=(2,7)$ to $(3,9)$ can be largely understood via insertion maps in the Cayley--Hamilton data.
The insertion of $\psi^2$ behaves essentially injectively and produces $11+11$ relations at $(3,9)$ from those at $(2,7)$.
In contrast, insertion of the fermionic letter $\partial\psi$ is more subtle:
\begin{itemize}
\item It is not injective, since inserting a second identical fermionic entry can lead to tautological zero.
For example, inserting $\partial\psi$ into $\PCH_{\partial\psi,\partial^3\psi}$ produces $\PCH_{\partial\psi,\partial\psi,\partial^3\psi}=0$.

\item It can split a single lower-rank relation into multiple inequivalent higher-rank relations depending on the insertion position.
For instance, while $\tr\psi\,\PCH_{\partial\psi\psi,\partial\psi\psi}$ and $\tr\psi\,\PCH_{\psi\partial\psi,\psi\partial\psi}$ are equivalent at rank $2$,
after inserting $\partial\psi$ they lead to inequivalent relations 
$\tr\psi\,\PCH_{\partial\psi,\partial\psi\psi,\partial\psi\psi}$ and $\tr\psi\,\PCH_{\partial\psi,\psi\partial\psi,\psi\partial\psi}$.
\end{itemize}
Empirically, the extra $4+4$ relations at $(3,9)$ beyond the $\psi^2$-image are precisely those generated by new nontrivial $\partial\psi$-insertions.
Moreover, starting at sufficiently large charge along the $(N\to N+1,\;n\to n+2)$ diagonal, the $\psi^2$ insertion dominates, and the diagonal stabilizes into Pattern~I, in agreement with Table~\ref{T:relations}.

\paragraph{Universality of Pattern I.} Motivated by this, we summarize our observations as the following conjecture:
\begin{conjecture}
Along each diagonal defined by $(N,n)\mapsto(N+1,n+2)$, the number of trace relations stabilizes to a constant sequence beginning at charge $n=3N$ (universality of Pattern~I). 
Moreover, for sufficiently large charge along the diagonal, all relations can be written in the form
\begin{equation}
\tr\psi\,\PCH_{\psi^2,\dots,\psi^2,X}\,,
\end{equation}
where $X$ represents a set of $N$ matrices and the set of $\tr\psi\,\PCH_X$ are precisely the trace relations at charge $3N$ and rank $N$.
\end{conjecture}
\ndt A heuristic explanation is that at large rank and relatively low charge (along the diagonals) one must reuse low-charge letters among the CH entries.
Since $\partial\psi$ cannot appear repeatedly (without trivializing the antisymmetrization), the repeated low charge entries are forced to be $\psi^2$, producing long $\psi^2$ strings.

We also note that the stabilized values along Pattern~I up to charge $20$ and rank $7$ are
$
\{1,\,4,\,15,\,52,\,166,\,511\}\,,
$
suggesting rapid (apparently exponential) growth of bosonic as well as fermionic operators. 
Of course, this growth is not visible in the index because of the cancellations due to bose-fermi pairing. 
It would be interesting to explore the holographic implication of this perfect cancellation between two exponentially growing sequences.

\bigskip

\subsubsection{Further thoughts}
While the pairing near the $n\lesssim 3N$ region admits a concrete description in terms of insertion maps, 
it remains mysterious why the pairing persists so cleanly when $n\gtrsim 3N$.
One possibly easier scenario is that at fixed rank $N$ and very large charge $n\gg 3N$, the CH data can be approximated by drawing the $N$ matrix inputs from an approximately i.i.d.\ distribution:
since the basic letters $\psi$ and $\partial$ both carry charge $1$, 
one expects roughly equal probability for a randomly chosen entry to be fermionic or bosonic.
Trading $\partial$ and $\psi$ then approximately induces a bijection between bosonic and fermionic relations.
Moreover, the product of $N$ i.i.d.\ random variables with equal bosonic/fermionic probability is itself equally likely to be bosonic or fermionic, 
leading to an approximate explanation of perfect pairing at large charge.
It would be interesting to quantify the error in these approximations and determine when (and how) pairing can fail.

This perspective suggests that to obtain a substantial imbalance between bosonic and fermionic trace relations---and thus potentially a large entropy growth at finite $N$---one should consider subsectors with at least \emph{one fermion}, and
\emph{more than one derivative} (or more generally, multiple distinct bosonic letters of the same charge).
This qualitative criterion is consistent with the general intuition advocated in \cite{Murthy:2020scj}. 
It would be valuable to formulate a precise condition for when finite-$N$ indices exhibit different types of asymptotic growth driven by trace-relation imbalances.


\section{Observed patterns in trace relations in different indices}
\label{sec:expindices}

In this short section we present some numerical patterns in the number of trace relations 
at rank~$N$ and charge~$n$. 
Recall from Section~\ref{sec:fermionplusderiv} that the~$\p \psi$ model is given by a single fermion~$\psi$ and 
a derivative~$\partial$, both with charge~$L=1$, and the corresponding single-letter index is given by $i^{\p \psi}(q) = -q/(1-q)$.
It is clear that all the single letters are fermionic in this model, and therefore the single-letter partition function 
is given by~$z^{\p \psi}(q)=-i^{\p \psi}(q)$.

Given the index~$I_N(q)$ and the partition function~$Z_N(q)$ for any~$N$, including~$N=\infty$, 
one can extract the generating functions for bosons and fermions separately as follows, 
\begin{equation}
\label{bosferm1}
    B_N(q)  \= \frac{1}{2} (Z_N(q)+I_N(q))\, , 
\qquad     F_N(q)  \= \frac{1}{2} (Z_N(q)-I_N(q))\, .
\end{equation}
We refer to the number of trace relations at a given~$N$ as the difference 
between the number of operators at infinite~$N$ and at a given~$N$ (for a fixed charge).

The first observation stems from the fact that, for any partition function arising from a single unitary 
matrix integral, the coefficients for a fixed charge stabilize as we increase the rank~\cite{Murthy:2020scj, Murthy:2022ien}. 
More precisely, as we cross a certain diagonal $n = a N + b$, where~$a$ and~$b$ are numbers that 
depend on the partition function in question, the coefficients equal the values at~$N=\infty$. 
In the model we study here, this phenomenon is seen in Tables~\ref{T:partition}, with~$a=2$, $b=1$.

Secondly, because the index is invariant as we vary the rank, the number of bosonic relations must be equal to 
the number of fermionic relations at every charge and rank. 
The number of such relations for the~$\p \psi$ model is given in Table~\ref{T:relations}.
Note that, as we decrease~$N$ from infinity, the diagonal~$n=2N+1$ is the first place where we begin to see the trace relations.

Further, there seems to be a non-trivial pattern obeyed by the number of these trace relations.
For each cell in such a table if we keep moving two cells down and one cell to the right we get 
what we will call an ``L-diagonal''.
If we observe the L-diagonals of Table~\ref{T:relations} (the number of trace relations for the index), 
we notice that the sequence of numbers along this diagonal eventually stabilizes into a constant sequence.

\smallskip

It turns out that there are similar patterns for all these indices we study, where either along 
L-diagonals or proper diagonals (i.e.~the next cell is one down and one right),  
the sequence of numbers seem to stabilize into either a constant sequence, or a linear sequence 
or sometimes even a quadratic sequence (where the second differences between terms are the same).
These patterns can be summarized as follows,
\begin{itemize}
    \item $\frac{1}{16}$-BPS index: L-diagonal pattern stabilizes into a quadratic sequence, 
    \item Schur index $\frac{1}{8}$-BPS: diagonal pattern stabilizes into a linear sequence,
    \item  \RI-BPS index $I^{\p \psi}$: L-diagonal pattern stabilizing into a constant sequence,
    \item $\frac{1}{2}$-BPS index: diagonal pattern stabilizes into a constant sequence, 
    \item The index $I^\psi$:  L-diagonal pattern stabilizes into a constant sequence. 
\end{itemize}
These data are presented in a GitHub repository,  see Appendix~\ref{app:supp}.
In general, the more supersymmetry one has, 
the lower the degree of the polynomial describing the general term of the sequence that the number of trace 
relations stabilizes to along diagonals (or L-diagonals). 
The patterns in~$I^\psi$ and in the~$\frac{1}{2}$-BPS index are relatively easy to understand 
due to their simple structure of single-matrix trace relations. 
We have seen in Section~\ref{sec:fermionplusderiv} that the patterns in the~\RI-BPS index~$I^{\p \psi}$ 
are already much more complicated 
because of the nature of multi-matrix trace relations. 
We leave the patterns in the Schur index and the~$\frac{1}{16}$-BPS index to future study.

\section*{Acknowledgements}

We thank  Chris Beem, Kasia Budzik, Yiming Chen, Stavros Garoufalidis, Zhengping Gui, Aidan Herderschee, Ji Hoon Lee, 
Mart\'i Rossell\'o, and Don Zagier for useful and enjoyable discussions and exchanges. 
S.~M.~acknowledges the support of the STFC UK grants ST/T000759/1, ST/X000753/1. 
Z.~J.~acknowledges the use of ChatGPT, an AI language model developed by OpenAI, for exploratory discussion, consistency checks, and reference checking.

\appendix

\section{Various limits of $\CN=4$ indices \label{app:Neq4indices}}

In this appendix we analyze different limits of the $\mathcal{N}=4$ superconformal index. 
We begin with a brief review of the $\mathcal{N}=4$ superconformal algebra in four dimensions 
and then show how the most general index is defined and computed. After this set-up, we then 
proceed to analyze the various limits of the index.

\subsection{$\mathcal{N}=4$ superconformal algebra in 4 dimensions}
\label{app:G}

In this subsection we give the full list of all non-zero relations of the $\mathcal{N}=4$ superconformal 
algebra in 4 dimensions, which is $psu(2,2|4)$. The bosonic sector of this algebra 
is~$so(2,4) \oplus su(4)$ and there are 16 $Q$-supercharges and 16 $S$-supercharges.
We use $\alpha, \beta = +,-$ for left-handed (undotted) spinor indices, $\dot{\alpha}, \dot{\beta} = +,-$ 
for right-handed (dotted) spinor indices, and $I,J=1,\dots,4$ for $su(4)_R$ indices throughout.
Throughout this appendix, Latin alphabet letters take the values $1, 2, 3, 4$ and Greek alphabet letters 
take the values $+, -$.

The generators of the algebra are 
\begin{itemize}
\item $J_{\alpha}^{\;\beta}$ and 
$\wt{J}_{\;\dot{\beta}}^{\dot{\alpha}}$\,, which are the generators of the two $SU(2)$ factors of the Lorentz transformations,
\item $P_{\alpha \dot{\alpha}}$\,, which is the generator of translations,
\item $H$\,, the generator of dilations,
\item $K^{\dot{\alpha} \alpha}$\,, which is the generator of special conformal transformations,
\item $R^I_{\; J}$\,, which is the generator of the $SU(4)$ $R$-symmetry, noting that it has $15$ independent components since it is traceless, i.e.~$\sum_{I=1}^4 R^I_{\;I}=0$\,,
\item $Q_{\; \alpha}^{I}\,, \wt{Q}_{I\dot{\alpha}}$ and their conjugates $S_I^{\;\alpha}\,, \wt{S}^{I\dot{\alpha}}$\,, which are the supercharges. 
\end{itemize}

The complete set of non-zero commutation/anti-commutation relations of the algebra are the following:

\smallskip
\ndt $\bullet$ Conformal algebra
\begin{equation}
\begin{split}
[J_{\alpha}^{\;\beta}, J_{\gamma}^{\;\delta}]&\=\delta_{\gamma}^{\;\beta}J_{\alpha}^{\;\delta}-\delta_{\alpha}^{\;\delta}J_{\gamma}^{\;\beta}\,,\quad [\wt{J}_{\;\dot{\beta}}^{\dot{\alpha}},\wt{J}_{\;\dot{\delta}}^{\dot{\gamma}}]\=\delta_{\;\dot{\delta}}^{\dot{\alpha}}\wt{J}_{\;\dot{\beta}}^{\dot{\gamma}}-\delta_{\;\dot{\beta}}^{\dot{\gamma}}\wt{J}_{\;\dot{\delta}}^{\dot{\alpha}}\,,\\
[J_{\alpha}^{\;\beta}, P_{\gamma \dot{\gamma}}]&\=\delta_{\gamma}^{\;\beta}P_{\alpha \dot{\gamma}}-\frac{1}{2}\delta_{\alpha}^{\;\beta}P_{\gamma \dot{\gamma}}\,,\quad [\wt{J}_{\;\dot{\beta}}^{\dot{\alpha}},P_{\gamma \dot{\gamma}}]\=\delta_{\;\dot{\gamma}}^{\dot{\alpha}}P_{\gamma \dot{\beta}}-\frac{1}{2}\delta_{\;\dot{\beta}}^{\dot{\alpha}}P_{\gamma \dot{\gamma}}\,,\\
[H,P_{\alpha \dot{\alpha}}]&\= P_{\alpha \dot{\alpha}}\,,\quad [H,K^{\dot{\alpha} \alpha}]\=-K^{\dot{\alpha} \alpha}\,,\\
[J_{\alpha}^{\;\beta}, K^{\dot{\gamma} \gamma}]&\=-\delta_{\alpha}^{\;\gamma}K^{\dot{\gamma} \beta}+\frac{1}{2}\delta_{\alpha}^{\;\beta}K^{\dot{\gamma} \gamma}\,,\quad [\wt{J}_{\;\dot{\beta}}^{\dot{\alpha}},K^{\dot{\gamma} \gamma}]\=-\delta_{\;\dot{\beta}}^{\dot{\gamma}}K^{\dot{\alpha} \gamma}+\frac{1}{2}\delta_{\;\dot{\beta}}^{\dot{\alpha}}K^{\dot{\gamma} \gamma}\,,\\
[K^{\dot{\alpha} \alpha},P_{\beta \dot{\beta}}]&\=\delta^{\dot{\alpha}}_{\;\dot{\beta}} J_\beta^{\;\alpha} + \delta_\beta^{\;\alpha} \wt{J}^{\dot{\alpha}}_{\;\dot{\beta}} + \delta_\beta^{\;\alpha} \delta^{\dot{\alpha}}_{\;\dot{\beta}} H\,,
\end{split}
\end{equation}

\smallskip
\ndt $\bullet$ Bosonic R-symmetry
\begin{equation}
[R^I_{\;J},R^K_{\;L}]\=\delta^K_{\;J}R^I_{\;L}-\delta^I_{\;L}R^K_{\;J}\,,
\end{equation}

\smallskip
\ndt $\bullet$ Supercharges
\begin{equation}
    \begin{split}
        \{Q_{\; \alpha}^{I},\wt{Q}_{J\dot{\alpha}} \}&\=\delta^I_{\;J}P_{\alpha \dot{\alpha}}\,,\quad \{S_I^{\;\alpha}, \wt{S}^{J\dot{\alpha}} \}\=\delta^J_{\;I}K^{\dot{\alpha}\alpha} \,,\\
\{Q_{\; \alpha}^{I},S_J^{\;\beta} \}&\=\delta^I_{\;J} J_\alpha^{\; \beta}-\delta_\alpha^{\; \beta} R^I_{\;J}+ \frac{1}{2} \delta_\alpha^{\; \beta} \delta^I_{\;J} H\,,\\
\{\wt{Q}_{I\dot{\alpha}}, \wt{S}^{J\dot{\beta}} \}&\=\delta^J_{\;I}\wt{J}^{\dot{\beta}}_{\; \dot{\alpha}}+\delta^{\dot{\beta}}_{\; \dot{\alpha}}R^J_{\;I}+\frac{1}{2}\delta^{\dot{\beta}}_{\; \dot{\alpha}}\delta^J_{\;I}H\,,
    \end{split}
\end{equation}

\smallskip
\ndt $\bullet$ Conformal transformations of the supercharges
\begin{equation}
\begin{split}
[J_{\alpha}^{\;\beta}, Q_{\; \gamma}^{I}]&\=\delta_{\gamma}^{\;\beta}Q_{\; \alpha}^{I}-\frac{1}{2}\delta_{\alpha}^{\;\beta}Q_{\; \gamma}^{I}\,,\quad [\wt{J}_{\;\dot{\beta}}^{\dot{\alpha}},\wt{Q}_{I\dot{\gamma}}]\=\delta_{\;\dot{\gamma}}^{\dot{\alpha}}\wt{Q}_{I\dot{\beta}}-\frac{1}{2}\delta_{\;\dot{\beta}}^{\dot{\alpha}}\wt{Q}_{I\dot{\gamma}}\,,\\
[J_{\alpha}^{\;\beta}, S^{\; \gamma}_{I}]&\=-\delta_{\alpha}^{\;\gamma}S^{\; \beta}_{I}+\frac{1}{2}\delta_{\alpha}^{\;\beta}S^{\; \gamma}_{I}\,,\quad [\wt{J}_{\;\dot{\beta}}^{\dot{\alpha}},\wt{S}^{I\dot{\gamma}}]\=-\delta_{\;\dot{\beta}}^{\dot{\gamma}}\wt{S}^{I\dot{\alpha}}+\frac{1}{2}\delta_{\;\dot{\beta}}^{\dot{\alpha}}\wt{S}^{I\dot{\gamma}}\,,\\
[P_{\alpha \dot{\alpha}},S^{\; \beta}_{I}]&\=-\delta_{\alpha}^{\; \beta} \wt{Q}_{I\dot{\alpha}}\,,\quad [P_{\alpha \dot{\alpha}},\wt{S}^{I\dot{\beta}}]\=-\delta^{\dot{\beta}}_{\; \dot{\alpha}}Q_{\; \alpha}^{I}\,,\\
[K^{\dot{\alpha} \alpha},Q_{\; \beta}^{I}]&\=\delta_\beta^{\; \alpha} \wt{S}^{I\dot{\alpha}}\,,\quad [K^{\dot{\alpha} \alpha},\wt{Q}_{I\dot{\beta}}]\=\delta^{\dot{\alpha}}_{\; \dot{\beta}}S^{\; \alpha}_{I}\,,\\
[H,Q_{\; \alpha}^{I}]&\=\frac{1}{2}Q_{\; \alpha}^{I}\,,\quad [H,\wt{Q}_{I\dot{\alpha}}]\=\frac{1}{2}\wt{Q}_{I\dot{\alpha}}\,,\quad [H,S_I^{\;\alpha}]\=-\frac{1}{2}S_I^{\;\alpha}\,,\quad [H,\wt{S}^{I\dot{\alpha}}]\=-\frac{1}{2}\wt{S}^{I\dot{\alpha}}\,,
\end{split}
\end{equation}

\smallskip
\ndt $\bullet$ R-symmetry transformations of the supercharges
\begin{equation}
    \begin{split}
        [R^I_{\;J},Q_{\; \alpha}^{K}]&\=\delta^K_{\;J}Q_{\; \alpha}^{I}-\frac{1}{4}\delta^I_{\;J}Q_{\; \alpha}^{K}\,,\quad [R^I_{\;J},\wt{Q}_{K\dot{\alpha}}]\=-\delta^I_{\;K}\wt{Q}_{J\dot{\alpha}}+\frac{1}{4}\delta^I_{\;J}\wt{Q}_{K\dot{\alpha}}\,,\\
[R^I_{\;J},S_K^{\;\alpha}]&\=-\delta^I_{\;K}S_J^{\;\alpha}+\frac{1}{4}\delta^I_{\;J}S_K^{\;\alpha}\,,\quad [R^I_{\;J},\wt{S}^{K\dot{\alpha}}]\=\delta^K_{\;J}\wt{S}^{I\dot{\alpha}}-\frac{1}{4}\delta^I_{\;J}\wt{S}^{K\dot{\alpha}}\,.
    \end{split}
\end{equation}
The bosonic part of the algebra has rank $3+3=6$ and therefore its Cartan subalgebra has dimension $6$.
The Cartan charges $(E,j,\wt{j},R_1\,,R_2\,,R_3)$ are chosen as follows
\begin{equation}
\label{Cartan}
\begin{split}
E &\= H, \qquad j\=J_+^{\; +}\= -J_-^{\; -}, \qquad  \wt{j}\=\wt{J}^{\dot{+}}_{\; \dot{+}}\= -\wt{J}^{\dot{-}}_{\; \dot{-}}\,,\\
R_1&\=R^1_{\;1}-R^2_{\;2}\,, \qquad R_2\=R^2_{\;2}-R^3_{\;3}\,, \qquad R_3\=R^3_{\;3}-R^4_{\;4}\,.
\end{split}
\end{equation}
The first three are the conformal Cartans, the scaling dimension and the two spins (angular momenta inside AdS$_5$), and the last three correspond to rotations in the three orthogonal planes inside the internal $\R^6$ defined by the six real scalars of $\mathcal{N}=4$ SYM (equivalently angular momenta inside S$^5$).

\subsection{Analysis and computation of limits of $\mathcal{N}=4$ index}

We begin by defining the most general index using the $\mathcal{N}=4$ superconformal algebra (without including any flavor/external symmetries). 
The first step is to choose a specific supercharge (and its conjugate) with which to define the index. We assume that this first choice can be done without loss of generality.
We will define our index with respect to $\Delta_{1 -}= E-2j-\frac{1}{2}(3R_1+2R_2+R_3)$, which means that the contributing letters will be those that are annihilated by $Q_{\; -}^{1}$ and $S_1^{\;-}$ and therefore have $\Delta_{1 -}=0$.

Next, in order to find all possible fugacities that we can include in the index, we compute the commutant 
of $Q_{\; -}^{1}$ and $S_1^{\;-}$ inside our superconformal algebra $psu(2,2|4)$. 
This commutant subalgebra is $psu(2,1|3)$ and has rank $4$.
We choose a basis for this commutant that is entirely made up of charges that 
only take non-negative values for any letter of our theory, since as we will later see this ensures 
that the limits we will analyze are all well-defined.
To do this we define the following (here the repeated indices are not summed over)
\begin{equation}
\label{deltas}
\begin{split}
    \Delta_{I \alpha}&\coloneqq 2\{Q_{\; \alpha}^{I},S_I^{\;\alpha} \}\=2J_\alpha^{\; \alpha}-2R^I_{\;I}+ H\,,\\ \quad \wt{\Delta}^I_{ \dot{\alpha}} &\coloneqq 2\{\wt{Q}_{I\dot{\alpha}}, \wt{S}^{I\dot{\alpha}} \}\=2\wt{J}^{\dot{\alpha}}_{\; \dot{\alpha}}+2R^I_{\;I}+H\,.
\end{split}
\end{equation}
Indeed, since these charges are anticommutators of supercharges with their corresponding conjugates, they are non-negative.

\begin{table}[h!]
\centering
\begin{tabular}{c|cccccc|c|m{4.5cm}} 
 \toprule
 $Q$ & $E$ & $j$ & $\wt{j}$ & $R_1$ & $R_2$ & $R_3$ & $\Delta$ & Commuting $\Delta$s \\ 
 \midrule
 $Q_{\; -}^{1}$ & $\frac{1}{2}$ & $-\frac{1}{2}$ & $0$ & $1$ & $0$ & $0$ & $E-2j-\frac{1}{2}(3R_1+2R_2+R_3)$ & \{$\Delta_{2+}$, $\Delta_{3+}$, $\Delta_{4+}$, $\wt{\Delta}^2_{ \dot{-}}$, $\wt{\Delta}^2_{ \dot{+}}$, $\wt{\Delta}^3_{ \dot{-}}$, $\wt{\Delta}^3_{ \dot{+}}$, $\wt{\Delta}^4_{ \dot{-}}$, $\wt{\Delta}^4_{ \dot{+}}$\} \\
 $Q_{\; +}^{1}$ & $\frac{1}{2}$ & $+\frac{1}{2}$ & $0$ & $1$ & $0$ & $0$ & $E+2j-\frac{1}{2}(3R_1+2R_2+R_3)$ & \{$\Delta_{2-}$, $\Delta_{3-}$, $\Delta_{4-}$, $\wt{\Delta}^2_{ \dot{-}}$, $\wt{\Delta}^2_{ \dot{+}}$, $\wt{\Delta}^3_{ \dot{-}}$, $\wt{\Delta}^3_{ \dot{+}}$, $\wt{\Delta}^4_{ \dot{-}}$, $\wt{\Delta}^4_{ \dot{+}}$\} \\
 $Q_{\; -}^{2}$ & $\frac{1}{2}$ & $-\frac{1}{2}$ & $0$ & $-1$ & $1$ & $0$ & $E-2j-\frac{1}{2}(-R_1+2R_2+R_3)$ & \{$\Delta_{1+}$, $\Delta_{3+}$, $\Delta_{4+}$, $\wt{\Delta}^1_{ \dot{-}}$, $\wt{\Delta}^1_{ \dot{+}}$, $\wt{\Delta}^3_{ \dot{-}}$, $\wt{\Delta}^3_{ \dot{+}}$, $\wt{\Delta}^4_{ \dot{-}}$, $\wt{\Delta}^4_{ \dot{+}}$\}\\
 $Q_{\; +}^{2}$ & $\frac{1}{2}$ & $+\frac{1}{2}$ & $0$ & $-1$ & $1$ & $0$ & $E+2j-\frac{1}{2}(-R_1+2R_2+R_3)$ & \{$\Delta_{1-}$, $\Delta_{3-}$, $\Delta_{4-}$, $\wt{\Delta}^1_{ \dot{-}}$, $\wt{\Delta}^1_{ \dot{+}}$, $\wt{\Delta}^3_{ \dot{-}}$, $\wt{\Delta}^3_{ \dot{+}}$, $\wt{\Delta}^4_{ \dot{-}}$, $\wt{\Delta}^4_{ \dot{+}}$\} \\
 $Q_{\; -}^{3}$ & $\frac{1}{2}$ & $-\frac{1}{2}$ & $0$ & $0$ & $-1$ & $1$ & $E-2j-\frac{1}{2}(-R_1-2R_2+R_3)$ & \{$\Delta_{1+}$, $\Delta_{2+}$, $\Delta_{4+}$, $\wt{\Delta}^1_{ \dot{-}}$, $\wt{\Delta}^1_{ \dot{+}}$, $\wt{\Delta}^2_{ \dot{-}}$, $\wt{\Delta}^2_{ \dot{+}}$,  $\wt{\Delta}^4_{ \dot{-}}$, $\wt{\Delta}^4_{ \dot{+}}$\} \\
 $Q_{\; +}^{3}$ & $\frac{1}{2}$ & $+\frac{1}{2}$ & $0$ & $0$ & $-1$ & $1$ & $E+2j-\frac{1}{2}(-R_1-2R_2+R_3)$ & \{$\Delta_{1-}$, $\Delta_{2-}$, $\Delta_{4-}$, $\wt{\Delta}^1_{ \dot{-}}$, $\wt{\Delta}^1_{ \dot{+}}$, $\wt{\Delta}^2_{ \dot{-}}$, $\wt{\Delta}^2_{ \dot{+}}$, $\wt{\Delta}^4_{ \dot{-}}$, $\wt{\Delta}^4_{ \dot{+}}$\} \\
 $Q_{\; -}^{4}$ & $\frac{1}{2}$ & $-\frac{1}{2}$ & $0$ & $0$ & $0$ & $-1$ & $E-2j-\frac{1}{2}(-R_1-2R_2-3R_3)$ & \{$\Delta_{1+}$, $\Delta_{2+}$, $\Delta_{3+}$, $\wt{\Delta}^1_{ \dot{-}}$, $\wt{\Delta}^1_{ \dot{+}}$, $\wt{\Delta}^2_{ \dot{-}}$, $\wt{\Delta}^2_{ \dot{+}}$,  $\wt{\Delta}^3_{ \dot{-}}$, $\wt{\Delta}^3_{ \dot{+}}$\} \\
 $Q_{\; +}^{4}$ & $\frac{1}{2}$ & $+\frac{1}{2}$ & $0$ & $0$ & $0$ & $-1$ & $E+2j-\frac{1}{2}(-R_1-2R_2-3R_3)$ & \{$\Delta_{1-}$, $\Delta_{2-}$, $\Delta_{3-}$, $\wt{\Delta}^1_{ \dot{-}}$, $\wt{\Delta}^1_{ \dot{+}}$, $\wt{\Delta}^2_{ \dot{-}}$, $\wt{\Delta}^2_{ \dot{+}}$, $\wt{\Delta}^3_{ \dot{-}}$, $\wt{\Delta}^3_{ \dot{+}}$\} \\
 $\wt{Q}_{1\dot{-}}$ & $\frac{1}{2}$ & $0$ & $-\frac{1}{2}$ & $-1$ & $0$ & $0$ & $E-2\wt{j}+\frac{1}{2}(3R_1+2R_2+R_3)$ & \{$\wt{\Delta}^2_{\dot{+}}$, $\wt{\Delta}^3_{\dot{+}}$, $\wt{\Delta}^4_{\dot{+}}$, $\Delta_{2 -}$, $\Delta_{2 +}$, $\Delta_{3 -}$, $\Delta_{3 +}$, $\Delta_{4 -}$, $\Delta_{4 +}$\} \\
 $\wt{Q}_{1\dot{+}}$ & $\frac{1}{2}$ & $0$ & $+\frac{1}{2}$ & $-1$ & $0$ & $0$ & $E+2\wt{j}+\frac{1}{2}(3R_1+2R_2+R_3)$ & \{$\wt{\Delta}^2_{\dot{-}}$, $\wt{\Delta}^3_{\dot{-}}$, $\wt{\Delta}^4_{\dot{-}}$, $\Delta_{2 -}$, $\Delta_{2 +}$, $\Delta_{3 -}$, $\Delta_{3 +}$, $\Delta_{4 -}$, $\Delta_{4 +}$\} \\
  $\wt{Q}_{2\dot{-}}$ & $\frac{1}{2}$ & $0$ & $-\frac{1}{2}$ & $1$ & $-1$ & $0$ & $E-2\wt{j}+\frac{1}{2}(-R_1+2R_2+R_3)$ & \{$\wt{\Delta}^1_{\dot{+}}$, $\wt{\Delta}^3_{\dot{+}}$, $\wt{\Delta}^4_{\dot{+}}$, $\Delta_{1 -}$, $\Delta_{1 +}$, $\Delta_{3 -}$, $\Delta_{3 +}$, $\Delta_{4 -}$, $\Delta_{4 +}$\} \\
 $\wt{Q}_{2\dot{+}}$ & $\frac{1}{2}$ & $0$ & $+\frac{1}{2}$ & $1$ & $-1$ & $0$ & $E+2\wt{j}+\frac{1}{2}(-R_1+2R_2+R_3)$ & \{$\wt{\Delta}^1_{\dot{-}}$, $\wt{\Delta}^3_{\dot{-}}$, $\wt{\Delta}^4_{\dot{-}}$, $\Delta_{1 -}$, $\Delta_{1 +}$, $\Delta_{3 -}$, $\Delta_{3 +}$, $\Delta_{4 -}$, $\Delta_{4 +}$\} \\
  $\wt{Q}_{3\dot{-}}$ & $\frac{1}{2}$ & $0$ & $-\frac{1}{2}$ & $0$ & $1$ & $-1$ & $E-2\wt{j}+\frac{1}{2}(-R_1-2R_2+R_3)$ & \{$\wt{\Delta}^1_{\dot{+}}$, $\wt{\Delta}^2_{\dot{+}}$, $\wt{\Delta}^4_{\dot{+}}$, $\Delta_{1 -}$, $\Delta_{1 +}$, $\Delta_{2 -}$, $\Delta_{2 +}$, $\Delta_{4 -}$, $\Delta_{4 +}$\} \\
 $\wt{Q}_{3\dot{+}}$ & $\frac{1}{2}$ & $0$ & $+\frac{1}{2}$ & $0$ & $1$ & $-1$ & $E+2\wt{j}+\frac{1}{2}(-R_1-2R_2+R_3)$ & \{$\wt{\Delta}^1_{\dot{-}}$, $\wt{\Delta}^2_{\dot{-}}$, $\wt{\Delta}^4_{\dot{-}}$, $\Delta_{1 -}$, $\Delta_{1 +}$, $\Delta_{2 -}$, $\Delta_{2 +}$, $\Delta_{4 -}$, $\Delta_{4 +}$\} \\
  $\wt{Q}_{4\dot{-}}$ & $\frac{1}{2}$ & $0$ & $-\frac{1}{2}$ & $0$ & $0$ & $1$ & $E-2\wt{j}+\frac{1}{2}(-R_1-2R_2-3R_3)$ & \{$\wt{\Delta}^1_{\dot{+}}$, $\wt{\Delta}^2_{\dot{+}}$, $\wt{\Delta}^3_{\dot{+}}$, $\Delta_{1 -}$, $\Delta_{1 +}$, $\Delta_{2 -}$, $\Delta_{2 +}$, $\Delta_{3 -}$, $\Delta_{3 +}$\} \\
 $\wt{Q}_{4\dot{+}}$ & $\frac{1}{2}$ & $0$ & $+\frac{1}{2}$ & $0$ & $0$ & $1$ & $E+2\wt{j}+\frac{1}{2}(-R_1-2R_2-3R_3)$ & \{$\wt{\Delta}^1_{\dot{-}}$, $\wt{\Delta}^2_{\dot{-}}$, $\wt{\Delta}^3_{\dot{-}}$, $\Delta_{1 -}$, $\Delta_{1 +}$, $\Delta_{2 -}$, $\Delta_{2 +}$, $\Delta_{3 -}$, $\Delta_{3 +}$\} \\
 \bottomrule
\end{tabular}
\caption{$Q$ supercharges for $\mathcal{N}=4$ SYM.}
\label{table:superdelta}
\end{table}

Using Appendix~\ref{app:G} we can write down the charges of all the $Q$ supercharges, 
calculate their corresponding $\Delta$ using \eqref{deltas} and find all $\Delta$s that commute with them. 
We summarize these in Table~\ref{table:superdelta} below:

From Table~\ref{table:superdelta} we can read off that the $\Delta$s that commute with $Q_{\; -}^{1}$, 
and thus also with $S_1^{\;-}$, are
\begin{equation}
\begin{split}
\Delta_{2 +}&\=E+2j-\frac{1}{2}(-R_1+2R_2+R_3)\,, \qquad 
\Delta_{3 +}\=E+2j-\frac{1}{2}(-R_1-2R_2+R_3)\,,\\
\Delta_{4 +}&\=E+2j-\frac{1}{2}(-R_1-2R_2-3R_3)\,, \vspace{0.2cm}\\ 
\wt{\Delta}^2_{ \dot{-}}&\=E-2\wt{j}+\frac{1}{2}(-R_1+2R_2+R_3)\,,\qquad 
\wt{\Delta}^2_{ \dot{+}}\=E+2\wt{j}+\frac{1}{2}(-R_1+2R_2+R_3)\,, \\
\wt{\Delta}^3_{ \dot{-}}&\=E-2\wt{j}+\frac{1}{2}(-R_1-2R_2+R_3) \,,\qquad 
\wt{\Delta}^3_{ \dot{+}}\=E+2\wt{j}+\frac{1}{2}(-R_1-2R_2+R_3)\,, \\
\wt{\Delta}^4_{ \dot{-}}&\=E-2\wt{j}+\frac{1}{2}(-R_1-2R_2-3R_3)\,,\qquad 
\wt{\Delta}^4_{ \dot{+}}\=E+2\wt{j}+\frac{1}{2}(-R_1-2R_2-3R_3) \,.
\end{split}
\end{equation}
Since the commutant subalgebra of $Q_{\; -}^{1}$ and $S_1^{\;-}$ has rank $4$, in order to choose a basis for it we need to choose 4 of the above $\Delta$s that are linearly independent. We will choose to define our index with fugacities that couple to $\wt{\Delta}^2_{ \dot{+}}, \wt{\Delta}^3_{ \dot{+}}, \wt{\Delta}^4_{ \dot{-}}, \wt{\Delta}^4_{ \dot{+}}$.
We could have also chosen to define our index with any other 4 of the above $\Delta$s that are linearly independent, and for each possible choice we could consider all possible limits of the fugacities going to zero in the resulting index.
By carefully considering all qualitatively different such combinations, and performing for each the same procedure as we will do here, we found that this particular choice of 4 $\Delta$s that we made here leads to an index whose limits give all possible indices that can be obtained from this procedure.
Therefore we will only just consider this one choice of $\Delta$s.

\bigskip

We denote the index as $I$ and the single-letter index as $i$, as in the rest of the paper. The most general index is defined as
\begin{equation}
\label{genindex}
I(a,b,c,d)\={\rm Tr}_{\mathcal{H}} \, (-1)^F \, e^{-\beta \Delta_{1-}}\,a^{\frac{1}{2} \wt{\Delta}^2_{\dot{+}} }\,b^{\frac{1}{2} \wt{\Delta}^3_{\dot{+}} }\,c^{\frac{1}{2} \wt{\Delta}^4_{\dot{-}} }\,d^{\frac{1}{2} \wt{\Delta}^4_{\dot{+}} }\,,
\end{equation}
where $a,b,c,d$ are the fugacities on which the index depends and the trace is taken over the entire Hilbert space of states of our theory.

We proceed by calculating the single-letter index. In order to do this, we need to know all the charges for all the contributing single-letters. We use Table~2 from \cite{Kinney:2005ej} to identify the contributing single-letters and their charges. We see that the letters that contribute to this index are 3 scalars, 5 fermions, 1 vector, 1 equation of motion and 2 derivatives. We adapt the table for our purposes and write the charges of the contributing letters in terms of $\wt{\Delta}^2_{ \dot{+}}, \wt{\Delta}^3_{ \dot{+}}, \wt{\Delta}^4_{ \dot{-}}, \wt{\Delta}^4_{ \dot{+}}$. We summarize these in Table~\ref{table:chletter} below:

\begin{table}[h!]
\centering
\begin{tabular}{m{3cm}|cccc}
 \toprule
 Letter & $\wt{\Delta}^2_{\dot{+}}$ & $\wt{\Delta}^3_{\dot{+}}$ & $\wt{\Delta}^4_{\dot{-}}$ & $\wt{\Delta}^4_{\dot{+}}$ \\ 
 \midrule
 $X$ & $2$ & $0$ & $0$ & $0$ \\
 $Y$ & $0$ & $2$ & $0$ & $0$ \\
 $Z$ & $0$ & $0$ & $2$ & $2$ \\
 $\psi_{+,0;-++}$ & $0$ & $2$ & $2$ & $2$ \\
  $\psi_{+,0;+-+}$ & $2$ & $0$ & $2$ & $2$ \\
  $\psi_{+,0;++-}$ & $2$ & $2$ & $0$ & $0$ \\
  $\psi_{0,+;+++}$ & $2$ & $2$ & $0$ & $2$ \\
 $\psi_{0,-;+++}$ &  $0$ & $0$ & $2$ & $0$ \\
 $F_{++}$ & $2$ & $2$ & $2$ & $2$ \\
 \midrule $\partial_{++}\psi_{0,-;+++}+\partial_{+-}\psi_{0,+;+++}=0$ & $2$ & $2$ & $2$ & $2$ \\
 \midrule
  $\partial_{++}$ & $2$ & $2$ & $0$ & $2$ \\
  $\partial_{+-}$ & $0$ & $0$ & $2$ & $0$ \\
 \bottomrule
\end{tabular}
\caption{Charges of letters with $\Delta_{1-}=0$ for $\mathcal{N}=4$ SYM.}
\label{table:chletter}
\end{table}

The states that contribute to this index are those that are annihilated by $Q_{\; -}^{1}$ and thus this index is a $\frac{1}{16}$-BPS index. We recall that the single-letter index is calculated by the same trace as for the full index but over the Hilbert space of single-letters. Thus, using Table~\ref{table:chletter}, we have that
\begin{equation}
i(a,b,c,d)\=\frac{a+b+cd-bcd-a c d-a b-abd-c+a b c d+a b c d}{(1-abd)(1-c)}\,.
\end{equation}
Setting $a=b=q^2$, $c=q^3$, $d=q^{-1}$, we get the unrefined and simplified familiar form for this
\begin{equation}
i(q)\=\frac{3q^2-3q^4-2q^3+2q^6}{(1-q^3)^2}\=1-\frac{(1-q^2)^3}{(1-q^3)^2}\,,
\end{equation}
which is indeed the single-letter index of the unrefined $\frac{1}{16}$-BPS index.

\bigskip

Our goal is to derive other indices by taking various limits as different combinations of fugacities go to zero of this starting general index.
Our approach is inspired by \cite{Gadde:2011uv}, in which the authors studied interesting limits of $\mathcal{N}=2$ indices.
The key is that the fugacities couple to anticommutators of supercharges. These charges that couple to the fugacities are always non-negative for all our letters, and therefore limits where the fugacities go to 0 are well-defined. 
When taking such limits, only letters that are uncharged under the corresponding $\Delta$s contribute, since all other contributions are set to 0. 
Therefore, only letters that are annihilated by the corresponding supercharges will contribute and this is a straightforward way of taking limits of the fugacities that lead to indices with more supersymmetry.

We proceed by analyzing limits of $\eqref{genindex}$ as various combinations of the fugacities go to zero. We considered all possible combinations of fugacities going to zero. Many of them produce the same indices and therefore here we only present those that produce non-repeat indices.

\begin{itemize}
\item \subsubsection*{\boldsymbol{$a\rightarrow 0$}}
We take $a\rightarrow 0$ in \eqref{genindex} and we see from Table~\ref{table:chletter} that
the only letters contributing are $Y$, $Z$, $\psi_{+,0;-++}$, $\psi_{0,-;+++}$, $\partial_{+-}$, which are annihilated by $Q_{\; -}^{1}$, $\wt{Q}_{2 \dot{+}}$, and thus the resulting index is $\frac{1}{8}$-BPS. 
We have
\begin{equation}
i(b,c,d)\=\frac{b+cd-bcd-c}{1-c}\,.
\end{equation}
Setting $b=q$, $c=q^2$, $d=q^{-1}$, we get the unrefined and simplified familiar form for this
\begin{equation}
i(q)\=\frac{2q-2q^2}{1-q^2}\=\frac{2q}{1+q}\,,
\end{equation}
which we recognize as the single-letter index of the Schur index.

\item \subsubsection*{\boldsymbol{$a,c\rightarrow 0$}}
We take $a,c\rightarrow 0$ in \eqref{genindex} and we see from Table~\ref{table:chletter} that 
the only letter contributing is $Y$, which is annihilated by $Q_{\; -}^{1}$, $Q_{\; +}^{1}$, $Q_{\; -}^{3}$, $Q_{\; +}^{3}$, $\wt{Q}_{2 \dot{-}}$, $\wt{Q}_{2 \dot{+}}$, $\wt{Q}_{4 \dot{-}}$, $\wt{Q}_{4 \dot{+}}$, and thus the resulting index is $\frac{1}{2}$-BPS. 
We have
\begin{equation}
i(b)\=b\,.
\end{equation}
For uniformity we simply rename $b=q$, and make the observation that this single-letter index corresponds to that of the familiar $\frac{1}{2}$-BPS index
\begin{equation}
i(q)\=q\,.
\end{equation}

\item \subsubsection*{\boldsymbol{$b,d\rightarrow 0$}}
We take $b,d\rightarrow 0$ in \eqref{genindex} and we see from Table~\ref{table:chletter} that
the only letters contributing are $X$, $\psi_{0,-;+++}$, $\partial_{+-}$, which are annihilated by $Q_{\; -}^{1}$, $\wt{Q}_{3 \dot{+}}$, $\wt{Q}_{4 \dot{+}}$, and thus the resulting index is \RI-BPS.
We have
\begin{equation}
i(a,c)\=\frac{a-c}{1-c}\,.
\end{equation}
For uniformity we rename $a=p, c=q$
\begin{equation}
\label{irrindex}
i(p,q)\=\frac{p-q}{1-q}\,.
\end{equation}

\item \subsubsection*{\boldsymbol{$c,d\rightarrow 0$}}
We take $c,d\rightarrow 0$ in \eqref{genindex} and we see from Table~\ref{table:chletter} that
the only letters contributing are $X$, $Y$, $\psi_{+,0;++-}$ which are annihilated by $Q_{\; -}^{1}$, $\wt{Q}_{4 \dot{-}}$, $\wt{Q}_{4 \dot{+}}$, and thus the resulting index is $\frac{3}{16}$-BPS, since the contributing letters are annihilated by 3 out of the 16 supercharge pairs.\footnote{In \cite{Gaiotto:2021xce}, this index is referred to as a $\frac{1}{4}$-BPS index. We thank Ji Hoon Lee for a discussion on this matter.}

We have
\begin{equation}
i(a,b)\=a+b-ab\,.
\end{equation}
Setting $a=b=q$, we get the unrefined and simplified form for this
\begin{equation}
i(q)\=2q-q^2\,.
\end{equation}

\item \subsubsection*{\boldsymbol{$a,b,d\rightarrow 0$}}
We take $a,b,d\rightarrow 0$ in \eqref{genindex} and we see from Table~\ref{table:chletter} that
the only letters contributing are $\psi_{0,-;+++}$, $\partial_{+-}$, which are annihilated by $Q_{\; -}^{1}$, $\wt{Q}_{2 \dot{+}}$, $\wt{Q}_{3 \dot{+}}$, $\wt{Q}_{4 \dot{+}}$, and thus the resulting index is $\frac{1}{4}$-BPS.
We have
\begin{equation}
i(c)\=\frac{-c}{1-c}\,.
\end{equation}
For uniformity we simply rename $c=q$
\begin{equation}
i(q)\=-\frac{q}{1-q}\,.
\end{equation}
This is the single-letter index of the~$\p \psi$ model that we study in detail in this paper, which leads to the rank-invariant index.
\end{itemize}

\section{Proof of uniqueness of rank-invariant index}\label{app:unique}

In this appendix we discuss the uniqueness of the single-particle index~$i(q)$ such that the full index~$I_N(q)$ is rank-invariant, i.e., independent of~$N$. 
(Here, and in the following, we refer to the index, but clearly, for this discussion, the details of whether we are considering the index or partition function are not really important.)
The function  
\begin{equation}\label{eq:sol1}
i(q) \= -\frac{q^m}{1-q^m}
\=-(q^m+q^{2m}+q^{3m}+\cdots) \,,\qquad  m\ge 1
\end{equation}
is rank-invariant.  
This was already discussed in the main text in Section~\ref{sec:fermionplusderiv} for~$m=1$, and the extension to any~$m$ is trivial.

Following the discussion in the main text, we consider the index to be a power series with no constant term, i.e., 
\begin{equation}
i(q)\=\sum_{n\ge1}a_n \,q^n \,.
\end{equation} 
As we show below, just the equality 
\begin{equation}\label{eq:Feqmain}
I_1(q) \=  I_\infty(q) 
\end{equation}
is enough to fix~$i(q)$ uniquely up to scalings of the sort~$q \mapsto q^m$, $m=1,2,3,\dots$. 

\medskip

It is convenient to define the function 
\begin{equation}\label{eq:hq}
h(q) := \; -\log(1-i(q)) 
\= i(q)+\frac{i(q)^2}{2}+\frac{i(q)^3}{3}+\cdots \,,
\end{equation}
Expressing~$h(q)$ as the series 
\begin{equation} 
h(q) \; \equiv \; \sum_{n\ge1}h_n \, q^n \,,
\end{equation}
the equation~\eqref{eq:hq} implies a triangular set of equations relating~$h_n$ and~$a_n$.
In particular, $h_n-a_n$ is determined by the lower coefficients~$a_j$, $j < n$.  
The equation~\eqref{eq:hq} also implies 
\be
h'(q)  \= i'(q) + h'(q) \, i(q) \,.
\ee
Comparing the coefficient of the $q^{n-1}$ terms on both sides,  we obtain 
\begin{equation} \label{hnanrel}
h_n \= a_n + \frac{1}{n}\sum_{k=1}^{n-1}k \, h_k \, a_{n-k}\,.
\end{equation}

Now we analyze the equality~\eqref{eq:Feqmain}. 
The formulas for the index at~$N=1$ and~$N=\infty$ are given in equation~\eqref{eq:I11} and \eqref{eq:Iinf1}, respectively. We consider the logarithms, 
\begin{equation}
\log I_1(q) \= \sum_{r\ge1}\frac{i(q^r)}{r} \,,
\qquad
\log I_\infty(q)\= \sum_{r\ge1}-\log(1-i(q^r)) \= \sum_{r \ge 1} h(q^r) \,.
\end{equation}
Upon comparing the coefficient of $q^N$ on the two sides,
we obtain 
\begin{equation}\label{eq:Feqcomp}
a_N+\frac1N\sum_{\substack{d\mid N\\ d<N}}d\,a_d
\=
h_N+\sum_{\substack{d\mid N\\ d<N}}h_d \,.
\end{equation}

Now we are ready to present the proof of uniqueness. From now on, we disregard the trivial solution~$i(q)=0$, 
and suppose $i(q)\neq0$. 
The first step is to show that the leading coefficient of $i(q)$ has to be $-1$. 
Let~$m$ be the smallest number with
$a_m\neq0$, so~$a_n=0$, $n<m$. 
Then, \eqref{eq:hq} shows that (i) $h_1, \dots h_{m-1}$ vanish, (ii)~$h_m=a_m$, and (iii)~$h_{2m}=a_{2m}+a_m^{2}/2$.
Combining these observations with~\eqref{eq:Feqcomp} with $N=2m$ leads to  
\begin{equation}
a_{2m}+\frac{a_m}{2} \= h_{2m}+h_m \=a_{2m}+\frac{a_m^2}{2}+a_m \quad 
\Longrightarrow \quad a_m=-1 \,.
\end{equation}

It remains to show there is no other solution apart from~\eqref{eq:sol1} that starts with $-q^m$. 
Suppose
$i(q)=\sum_{n\ge1}a_n \,q^n$ and $\wt i(q)=\sum_{n\ge1} \wt a_n \, q^n$ are two 
different solutions to our problem that start with $-q^m$ and first differ at degree $j>m$, i.e.,
\begin{equation}\label{eq:k1}
    a_k\= \widetilde a_k \,,  \qquad k < j \,.
\end{equation}
Each series gives rise to corresponding series~$h(q)$, $\wt h(q)$. 
The triangular nature of the equations for the coefficients of~\eqref{eq:hq} implies the corresponding
coefficients of~$h$ and~$\wt h$ also agree, i.e.,
\begin{equation}\label{eq:k2}
     h_k\= \widetilde h_k \,, \qquad k < j \,.
\end{equation}
Moreover, because of the equation~\eqref{hnanrel},
we have 
\begin{equation}
h_j-\widetilde h_j\=a_j-\widetilde a_j \,.
\end{equation}

Now compare the two sides of the equation \eqref{eq:Feqcomp} for $i(q)$ at degree~$m+j$.
We write the difference (which, of course, vanishes) as
\begin{equation}\label{eq:FD}
    0 \= D \; \equiv \; \Big(a_{m+j}+\frac{1}{m+j}\sum_{\substack{d\mid (m+j)\\ d<m+j}}d\,a_d\Big) \; - \;\Big(h_{m+j}+\sum_{\substack{d\mid (m+j)\\ d<m+j}}h_d\Big) \,.
\end{equation}
There is a similar equation for~$\widetilde i(q)$, and we can now compare the two equations. 
In $D-\wt D$, all the proper-divisor terms in \eqref{eq:FD} cancel, since every proper divisor~$d$ of~$m+j$ is
less than~$j$. 
So we have
\begin{equation}\label{eq:FDD}
   0 \= D-\wt D\=\left(a_{m+j}-h_{m+j}\right)-\left(\wt a_{m+j}-\wt h_{m+j}\right) \,.
\end{equation}
Now we use \eqref{hnanrel} to arrive at
\begin{equation}
\begin{aligned}
       & \left(a_{m+j}-h_{m+j}\right)-\left(\wt a_{m+j}-\wt h_{m+j}\right)\\
       & \quad \=-\frac{1}{m+j}\sum_{k=1}^{m+j-1}k \, h_k \, a_{m+j-k} \,+\,  \frac{1}{m+j}\sum_{k=1}^{m+j-1}k \, \wt h_k \, \wt a_{m+j-k} \,.
\end{aligned}
\end{equation}
In this sum, the terms with $1\leq k<m$ vanish because $h_{k}=0$; the terms with $m<k<j$ cancel between $D$ and $\wt D$ because of~\eqref{eq:k1}, \eqref{eq:k2}; 
and the terms with $j<k\leq m+j-1$ also vanish because $a_{m+j-k}=0$. 
Thus, the only remaining terms are 
\begin{equation}
0 \= D-\wt D\=\frac{m }{m+j}(h_m \, a_j-\wt h_m \, \widetilde a_j)
+
\frac{j }{m+j}(h_j \, a_m -\widetilde h_j \, \wt a_m ) \,.
\end{equation}
Using~$a_m=h_m=-1$, $h_j-\widetilde h_j\=a_j-\widetilde a_j$,  we obtain
\begin{equation}
0 \= a_j-\widetilde a_j \,,
\end{equation}
contradicting the choice of $j$. 

Therefore $i(q)=\widetilde i(q)$, that is to say the only non-trivial solutions  of~\eqref{eq:Feqmain} are 
\begin{equation} 
i(q) \= -\frac{q^m}{1-q^m}\quad m  \in \mathbb{N} \,.
\end{equation}

\section{Special function identities following from rank invariance \label{app:identities}}

Starting from the identity~\eqref{eq:INdiff}, 
we can obtain interesting identities among special functions as corollaries, 
simply by considering the coefficient of each power of~$q$. 
The rank invariance of the index of the $\p \psi$
can be stated as the following identity
for~$N=2,3,\dots$, 
\begin{equation}
\label{eq:INdiff1}
   I^{\p \psi}_{N}(q) - I^{\p \psi}_{N-1}(q) 
   \= \sum_{\boldsymbol\lambda}  i^{\p \psi}_{\boldsymbol\lambda} (q) \,\kappa_N(\boldsymbol\lambda) \=0 \,,
\end{equation}
\begin{equation}
    \kappa_N(\boldsymbol\lambda) \= 
     \frac{1}{z_{\boldsymbol\lambda}} \sum_{\ell(\boldsymbol\mu)=N} \chi^{\boldsymbol\mu} (\boldsymbol\lambda)^2 \,,
\end{equation}
with the single-letter index given by~$i^{\p \psi}(q)=-q/(1-q)$. 
Recall that~$\kappa_N$ vanishes when $|\boldsymbol\lambda|<N$, so that the sum 
in~\eqref{eq:INdiff1} starts at~$\boldsymbol\lambda \ge N$. 
We use the notations given below~\eqref{eq:char_exp1}. 
In particular, $\boldsymbol\lambda = \prod_{j\ge1} (j)^{k_j}$, $|\boldsymbol\lambda|=\sum_j j k_j$, $\ell(\boldsymbol\lambda)=\sum_j k_j$,
and~$i_{\boldsymbol\lambda}(q)= \prod_{j} i(q^{j})^{k_j}$.

Define~$P_n(\boldsymbol\lambda)$ through the following $q$-series, 
\begin{equation}
    \prod_{j}\left(\frac{1}{1-q^j}\right)^{k_j}\=\sum_{n=0}^\infty P_n(\boldsymbol\lambda)\,q^n \,.
\end{equation}
We can write~$P_n$ in terms of other special functions. In order to do this, recall Fa\`a di Bruno's formula on composite function derivatives, 
\begin{equation}
    {d^{n} \over dx^{n}}f(g(x))\=\sum _{k=1}^{n}f^{(k)}(g(x))\cdot B_{n,k}\left(g'(x),g''(x),\dots ,g^{(n-k+1)}(x)\right) \,,
\end{equation}
where $B_{n,k}$ is the exponential Bell polynomial. With~$f=\exp$ and $g(q)=\sum_{j}\log\left(\frac{1}{1-q^j}\right)^{k_j}$, we obtain 
\begin{equation}
    P_{n}(\boldsymbol\lambda)  \= \frac{1}{n!} \, \sum_{k=1}^{n} B_{n,k} \bigl(T_{\boldsymbol\lambda}(1),\ldots,T_{\boldsymbol\lambda}(n+1-k) \bigr) \, , \qquad   T_{\boldsymbol\lambda} (m) \coloneqq m! \, \sum_{ij=m} \frac{k_j}{i} \, . 
\end{equation}
Yet another way is to use the Taylor expansion of $\left(\frac{1}{1-q^j}\right)^{k_j}$ to obtain 
\begin{equation}    P_n({\boldsymbol\lambda})\=\sum_{\sum_jjn_j=n}\prod_j\frac{\Gamma(k_j+n_j)}{\Gamma(k_j)\Gamma(1+n_j)} \,.
\end{equation}

Now we could use the identity \ref{eq:INdiff1} to write down an identity for $P_n(\boldsymbol{\lambda})$. Recall in our one fermion one derivative model~$i^{\p \psi}(q)=-q/(1-q)$,
we have that $i^{\p \psi}_{\boldsymbol\lambda} (q) = O(q^{|\boldsymbol\lambda|})$. 
Then the coefficient of the $q^{N+m}$ term of the identity~\eqref{eq:INdiff1} can be written as,
for every~$N=2,3,\dots$
\begin{equation}
    \sum_{r=0}^m \sum_{{\boldsymbol\lambda} \, \vdash N+r} (-1)^{\ell({\boldsymbol\lambda})} \, P_{m-r} ({\boldsymbol\lambda}) \, \kappa_N ({\boldsymbol\lambda}) \= 0, \qquad m \= 1,2,3,\dots \,.
\end{equation}

\section{$I^{\p \psi}_2(q)=(q;q)_\infty$ and special function identities \label{app:Neq2index}}

Recall from~\eqref{eq:intid1} that the index~$I^{\p \psi}_N$ for~$N=2$ is given by
\begin{equation}
I^{\p \psi}_2(q)\= \frac{(q;q)_{\infty}^2}{2} \int_0^1 \int_0^1 du_1 \, du_2 \, 
(\textbf{e}(u_{12});q)_{\infty} \, (\textbf{e}(u_{21});q)_{\infty} \, .
\end{equation}
(Here we have used $x_i = \textbf{e}(u_{i})=e^{2 \pi i \, u_i}$ and the notation $u_{ij}=u_i-u_j$.
Recall also that the $q$-Pochhammer symbol is given by~$(x;q)_n = \prod_{i=0}^{n-1} (1-x \, q^i)$.)
Using the following $q$-Pochhammer identity
\begin{equation}
    (x;q)_\infty\= \sum_{m=0}^\infty \, (-1)^{m} \, \frac{q^{\binom{m}{2}}}{(q;q)_m}  \, x^m \,,  
\end{equation}
we can express the integral as follows,
\begin{equation}
\begin{split}
I^{\p \psi}_2(q) 
&\= \frac{(q;q)_{\infty}^2}{2} \int_0^1 \int_0^1 du_1 \, du_2 \sum_{n,m=0}^{\infty} 
\frac{(-1)^{n+m} \, q^{\binom{n}{2}+\binom{m}{2}}}{(q;q)_n \, (q;q)_m} \,  \textbf{e}((n -m)u_{12}) \\
&\= \frac{(q;q)_{\infty}^2}{2} \, \sum_{n=0}^{\infty} \frac{q^{n(n-1)}}{(q;q)_n^2} \,.
\end{split}
\end{equation}
Therefore, the statement $I^{\p \psi}_2(q)=(q;q)_\infty$ is equivalent to the identity 
\begin{equation} \label{eq:I2equiv}
\frac{1}{(q;q)_{\infty}}\= \sum_{n=0}^{\infty} \frac{q^{n(n-1)}}{2\,(q;q)_n^2}\,.
\end{equation}

In order to proceed, we will use the result from~\cite{Andrews_Eriksson_2004} (Chapter~8, Exercise~101), which states that
\begin{equation}
\label{math1}
    \sum_{j=0}^N \binom{N}{j}_{q} \,\, \frac{z^j \, q^{j^2}}{(1-zq)(1-zq^2)\cdots (1-zq^j)}\= \prod_{n=1}^N \frac{1}{(1-zq^n)} \, ,
\end{equation}
where the so-called \emph{$q$-binomial coefficient} is defined as 
\begin{equation}
    \binom{N}{j}_{q} \= \frac{(q;q)_N}{(q;q)_j \, (q;q)_{N-j}}\,.
\end{equation}

In particular, setting $z=q$, taking the limit $N\rightarrow \infty$ in \eqref{math1}, and
multiplying both sides with $\frac{1}{1-q}$, we get
\begin{equation}
    \frac{1}{(q;q)_\infty}\= \frac{1}{1-q}+\sum_{n=1}^\infty \frac{q^{n(n+1)}}{(q;q)_{n+1} \, (q;q)_{n}}\,.
\end{equation}
We can now manipulate the expression as follows, 
\begin{equation}
\label{inbet}
\begin{split}
    \frac{1}{(q;q)_\infty} 
    & \=\frac{1}{1-q}+\sum_{k=2}^\infty \frac{q^{k(k-1)} \,(1-q^k)}{(q;q)_{k}^2}
    \=\frac{1}{1-q}+\sum_{k=2}^\infty \frac{q^{k(k-1)} }{(q;q)_{k}^2}-\sum_{k=2}^\infty \frac{q^{k^2} }{(q;q)_{k}^2}\,,\\
    &\= \sum_{k=0}^\infty \frac{q^{k(k-1)} }{(q;q)_{k}^2}-\sum_{k=0}^\infty \frac{q^{k^2} }{(q;q)_{k}^2} \,.
\end{split}    
\end{equation}
Similarly, setting $z=1$ in~\eqref{math1} and following a similar sequence of steps, 
we obtain\footnote{This is also the result (\cite{Andrews_Eriksson_2004}, Chapter 8, (8.2)).}  
\begin{equation}
     \frac{1}{(q;q)_\infty}\=\sum_{k=0}^\infty \frac{q^{k^2} }{(q;q)_{k}^2} \, .
\end{equation}
Upon combining the last two expressions, we obtain the identity~\eqref{eq:I2equiv} that we set out to prove. 

\section{Connection with the double-scaled SYK model\label{app:DSSYK}}
In this section, we point out an intriguing relation between the index of
the $\partial\psi$ theory and the infinite-temperature partition function of
the double-scaled SYK (DSSYK) model
\cite{Berkooz:2018qkz,Berkooz:2018jqr}, in the normalization used in
\cite{Gaiotto:2024kze}. More precisely, using the results obtained for the index of
the $\partial\psi$ theory, we show that an observable
constructed from the DSSYK energy function $E(\theta)$ is simply related to
the partition function.

This relation can be seen most clearly via the integral representation~\eqref{eq:intIN1} of the index with~$N=2$,
\begin{equation}
    \begin{aligned}
        I_2^{\partial\psi}(q)&=\frac{1}{2}\oint_{\mathbb{T}^2}\frac{dx_1}{2\pi x_1}\frac{dx_2}{2\pi x_2}(q;q)_\infty^2(x_1x_2^{-1};q)_\infty(x_2x_1^{-1};q)_\infty\\
        &= \frac{1}{8\pi^2}\int_0^{2\pi}d\theta_1\int_0^{2\pi}d\theta_2\,(q;q)_\infty^2(e^{i(\theta_1-\theta_2)};q)_\infty(e^{i(\theta_2-\theta_1)};q)_\infty\\
        &= \frac{1}{4\pi^2}  \int d\wt \theta d\theta\,(q;q)_\infty^2(e^{2i\theta};q)_\infty(e^{-2i\theta};q)_\infty\\
    \end{aligned}
\end{equation}
where we define $x_1=e^{i\theta_1}$, $x_2=e^{i\theta_2}$, $\wt \theta=\frac{\theta_1+\theta_2}{2}$, and $\theta=\frac{\theta_1-\theta_2}{2}$.
Now, since that the integrand is independent of $\wt\theta$, we can integrate it out to obtain
\begin{equation}
    \begin{aligned}
        & I_2^{\partial\psi}(q) \\
        &\= \frac{1}{4\pi^2} \left(\int_{-\pi}^{\pi} d\theta \int_{\max{\theta,-\theta}}^{\min{2\pi-\theta,2\pi+\theta}} d\theta\right)\,(q;q)_\infty^2 (e^{2i\theta};q)_\infty(e^{-2i\theta};q)_\infty\\
        &\=\frac{1}{4\pi^2} \left(\int_{-\pi}^{0} d\theta \int_{-\theta}^{2\pi+\theta} d\theta + \int_{0}^{\pi} d\theta \int_{\theta}^{2\pi-\theta} d\theta\right)\,(q;q)_\infty^2 (e^{2i\theta};q)_\infty(e^{-2i\theta};q)_\infty\\
        &\=\frac{1}{4\pi^2} \left(\int_{-\pi}^{0} d\theta (2\pi+2\theta) + \int_{0}^{\pi} d\theta (2\pi-2\theta)\right)\,(q;q)_\infty^2 (e^{2i\theta};q)_\infty(e^{-2i\theta};q)_\infty\\
        &\= (q;q)_\infty\int_{0}^{\pi} \frac{d\theta}{2\pi^2}  (2\pi-2\theta)\,(q;q)_\infty(e^{2i\theta};q)_\infty(e^{-2i\theta};q)_\infty\\
        &\= (q;q)_\infty\int_{0}^{\pi} \frac{d\theta}{2\pi}  (q;q)_\infty(e^{2i\theta};q)_\infty(e^{-2i\theta};q)_\infty \,.
    \end{aligned}
\end{equation}
The last equality in the above equation is obtained using the $\theta\to\pi-\theta$ symmetry of the integrand $(q;q)_\infty(e^{2i\theta};q)_\infty(e^{-2i\theta};q)_\infty$. The fact that the index for any $N$ is $I_N^{\partial\psi}(q)=(q;q)_\infty$ implies that the integral of $\theta$ in the above equation 
should be equal to~1:
\begin{equation}\label{eq:curious}
    \int_{0}^{\pi} \frac{d\theta}{2\pi}  (q;q)_\infty(e^{2i\theta};q)_\infty(e^{-2i\theta};q)_\infty = 1\,,
\end{equation}
which can also be checked straightforwardly term by term in the $q$-series (see appendix \ref{app:Neq2index}).

Now the relation to DSSYK begins with the observation that 
the integral in the equation above is very similar to that of the DSSYK partition function at infinite temperature (in the normalization of \cite{Gaiotto:2024kze}) with the identification $q=e^{-\lambda}$ where $\lambda$ is the double-scaling coupling, see Equation~(1.5) in~\cite{Gaiotto:2024kze}, which states that
\begin{equation}\label{eq:DSSYK}
    Z_\text{DSSYK}(q)\= \int_0^\pi\frac{d\theta}{\pi} \, (q;q)'_\infty \, (e^{2i\theta};q)'_\infty \, (e^{-2i\theta};q)'_\infty \,.
\end{equation}
Note that the definition of $q$-Pochammer symbol in \cite{Gaiotto:2024kze} is different from our convention.
Their definition is~$(a;q)'_\infty \equiv\prod_{k=1}^\infty(1-a q^k)$, 
while ours is $(a;q)_\infty \equiv  \prod_{k=0}^\infty(1-a q^k)$.
The $N=2$ $\partial\psi$ index can be written in their notation as
\begin{equation}
    \begin{aligned}
        I_2^{\partial\psi}(q)&=\frac{1}{2} (q;q)_\infty\int_{0}^{\pi} \frac{d\theta}{\pi} (q;q)_\infty(e^{2i\theta};q)_\infty(e^{-2i\theta};q)_\infty\\
        &=\frac{1}{2} (q;q)'_\infty\int_{0}^{\pi} \frac{d\theta}{\pi} (1-q)^24\sin^2\theta(q;q)'_\infty(e^{2i\theta};q)'_\infty(e^{-2i\theta};q)'_\infty \,.
    \end{aligned}
\end{equation}
The integral in \eqref{eq:curious} can be written in their notation as
\begin{equation}
    \begin{aligned}
      2 =& \int_{0}^{\pi} \frac{d\theta}{\pi} (1-q)4\sin^2\theta(q;q)'_\infty(e^{2i\theta};q)'_\infty(e^{-2i\theta};q)'_\infty\\
      =& \int_{0}^{\pi} \frac{d\theta}{\pi} (1-q)\big(4-\lambda(1-q)E(\theta)^2\big)(q;q)'_\infty(e^{2i\theta};q)'_\infty(e^{-2i\theta};q)'_\infty
    \end{aligned}
\end{equation}
The integral in the second line above can be seen as the expectation value of 
some insertion in the DSSYK model where $E(\theta)=\frac{-2\cos\theta}{\sqrt{\lambda(1-q)}}$ is the energy in DSSYK and $\lambda$ is the double scaling coupling of DSSYK. 
After some small manipulations, we arrive at a relation between the normalized expectation value of the squared energy and the partition function, 
\begin{equation}
    \langle E^2\rangle_\text{DSSYK} \= \frac{4}{\lambda(1-q)} \biggl( 1-\frac{1}{2(1-q)Z_\text{DSSYK}} \biggr) \,.
\end{equation}
See \cite{Gaiotto:2024kze} for the concrete relation between
\eqref{eq:DSSYK} and the Schur half-index of $SU(2)$ Seiberg-Witten theory.

\section{Supplementary data}\label{app:supp}
In this appendix, we give tables with the number of bosonic and fermionic operators and with the 
number of bosonic and fermionic trace relations contributing to various indices. This follows from 
the discussion in Section~\ref{sec:expindices}, where we gave these tables for the $\frac{1}{16}$-BPS, Schur ($\frac{1}{8}$-BPS), $\frac{1}{2}$-BPS 
and one-fermion model indices.
Note that no Fermionic operators (and thus also no corresponding Fermionic trace relations) contribute 
to the $\frac{1}{2}$-BPS index at any rank and charge so the corresponding tables are trivial (filled with 
just zeroes) and are thus omitted. 
We also share the raw data and code used in Section~\ref{sec:polarizedCH}.
The data is available on \href{https://github.com/j6z6m6/FermionTraceRelations}{https://github.com/j6z6m6/FermionTraceRelations}.

\bibliography{FTR}
\bibliographystyle{JHEP}

\end{document}